\documentclass[sigconf,authorversion]{acmart}
\usepackage[T1]{fontenc}
\def\BibTeX{{\rm B\kern-.05em{\sc i\kern-.025em b}\kern-.08emT\kern-.1667em\lower.7ex\hbox{E}\kern-.125emX}}
\acmConference[SCA 2019]{SCA: Symposium on Computer Animation}{July 26--28, 2019}{Los Angeles, CA}

\usepackage{amsmath,amsfonts,amssymb}
\usepackage{overpic}
\usepackage{multirow}
\usepackage{wrapfig}

\newcommand{\myreffig}[1]{Fig.~\ref{#1}}
\newcommand{\myrefeq}[1]{Eq.~\eqref{#1}}
\newcommand{\myreftab}[1]{Table~\ref{#1}}
\newcommand{\myrefsec}[1]{Sec.~\ref{#1}}

\newcommand{\myavg}{\sum}
\newcommand{\resizeEq}[3]{\begin{equation}%equations
\resizebox{#3\hsize}{!}{%
$\begin{array}{@{}ll}
\begin{aligned}
#1
\end{aligned}
\end{array}$}  \label{#2}
\end{equation}}

\definecolor{purple}{rgb}{0.36,0.0,0.6}
\definecolor{green}{rgb}{0.0,0.4,0.0}
\definecolor{orange}{rgb}{1,0.5,0}
\definecolor{pink}{rgb}{0.858, 0.188, 0.478}
\definecolor{LightCyan}{rgb}{0.8,0.86,1}
\definecolor{darkgrey}{rgb}{0.5,0.5,0.5}
\definecolor{nilscol}{rgb}{0.2,0.15,0.8}
\definecolor{nilsNewcol}{rgb}{0.9,0.1,0.0}

\newcommand{\MaxTodo}[1]{{#1}}
\newcommand{\MaxTodoNew}[1]{{#1}}

\newcommand{\YouAdd}[1]{{#1}}

\newcommand{\marie}[1]{{#1}}
  
\newcommand{\nilsNew}[1]{{#1}}  
\newcommand{\nils}[1]{{#1}}

\begin{document}

\title{A Multi-Pass GAN for Fluid Flow Super-Resolution}

\author{Maximilian Werhahn}
\affiliation{
  \institution{Technical University of Munich}
}
\email{maxi.werhahn@gmx.de}

\author{You Xie}
\affiliation{
  \institution{Technical University of Munich}
}
\email{you.xie@tum.de}

\author{Mengyu Chu}
\affiliation{
  \institution{Technical University of Munich}
}
\email{mengyu.chu@tum.de}

\author{Nils Thuerey}
\affiliation{
 \institution{Technical University of Munich}
}
\email{nils.thuerey@tum.de}

\begin{abstract}
We propose a novel method to up-sample volumetric functions with generative neural networks using several orthogonal passes. Our method decomposes generative problems on Cartesian field functions into multiple smaller sub-problems that can be learned more \marie{efficiently}. Specifically, we utilize two separate generative adversarial networks: the first one up-scales slices which are \marie{parallel to} the $XY$-plane, whereas the second one refines the whole volume along the $Z-$axis working on slices in the $YZ$-plane. In this way, we \marie{obtain} full coverage for the 3D target function and can leverage spatio-temporal supervision with a set of discriminators. Additionally, we demonstrate that our method can be combined with curriculum learning and progressive growing approaches. We arrive at a first method that can up-sample volumes by a factor of eight along each dimension, i.e., increasing the number of degrees of freedom by 512. Large volumetric up-scaling factors such as this one \marie{have previously not been attainable} as the required number of weights in the neural networks \marie{renders} adversarial training runs prohibitively difficult. We demonstrate the generality of our trained networks with a series of comparisons to previous work, a variety of complex 3D results, and an analysis of the resulting performance.
\end{abstract}




\begin{teaserfigure}
\centering
  \vspace{-12pt}
  \includegraphics[width=0.997\textwidth]{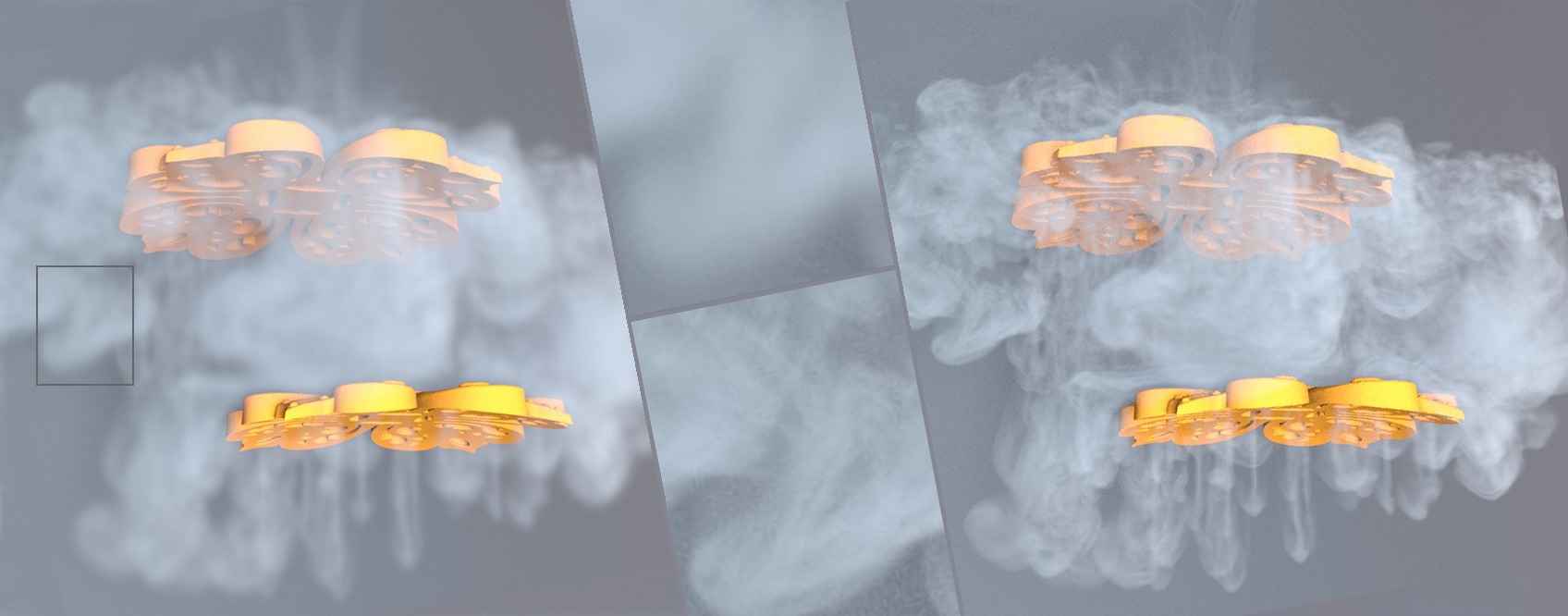}
  \vspace{-16pt}
  \caption{\small {%An 
  A $\mathbf{100^3}$ simulation (left) is up-sampled with our multi-pass GAN \marie{by a factor of $8$} to a resolution of $\mathbf{800^3}$ (right). The generated volume contains more than $\mathbf{500}$ million cells for every time step of the simulation. 
  \marie{In the middle inset, the left box is repeated as zoom-in for both resolutions.}}}
  \Description{Teaser}
  \label{fig:teaser}
\end{teaserfigure}
\vspace{2pt}
\maketitle

\section{Introduction}
\YouAdd{Deep learning\marie{-}based generative models, in particular generative adversarial networks (GANs) \cite{goodfellow2014generative}, are widely used for synthesis\marie{-}related learning tasks. GANs contain a generator, which can be trained to achieve a specific task, and a discriminator network that efficiently represents a learned loss function. The discriminator drives the generated data distribution to be close to the reference data distribution.} There is no explicit content-based loss function for training the generator, \YouAdd{and as a consequence}\marie{,} GANs perform very well for problems with multiple solutions, i.e.\marie{,} multimodal problems, such as super-resolution (SR) tasks. Most of the generative models focus on \marie{two-dimensional (2D)} data, such as images, as 
processing higher dimensional data quickly becomes prohibitively expensive.
However, \marie{as} our world is full of \marie{three-dimensional (3D)} objects, \YouAdd{learning} 3D relationships is an \marie{important and challenging task}. In the following, we will propose a volumetric training pipeline called {\em multi-pass GAN}. Our method breaks down the generative \marie{task} to infer large Cartesian density functions in space and time into multiple orthogonal passes, each of which represents a much more manageable inference problem.

We target a fluid flow setting where the multi-pass GAN solely uses 2D slices for training but \marie{is able to} learn 3D relationships by processing the data multiple times.
This is especially beneficial for training: finding a stable minimum for the coupled non-linear optimization of a GAN training grows in complexity with the number of variables to train. Our multi-pass network significantly reduce\marie{s} the number of variables and in this way stabilize\marie{s} the training, which in turn makes it possible to train networks with complexities that would be \marie{i}nfeasible with traditional approaches.
Our approach can potentially be used in a variety of 3D training cases, such as 3D data SR, classification\marie{,} or data synthesis.
In the following, we demonstrate its capabilities for fluid flow SR\MaxTodoNew{, more specifically for buoyant smoke \marie{SR},} in conjunction with the tempoGAN~\cite{xie2018tempogan} and the progressive growing of GANs~\cite{wang2018progressive} architectures. 

For SR problems, there are inherent challenges. First \marie{of all}, most SR algorithms focus on $4 \times$ 2D data SR.
The large size of 3D volumetric data makes it difficult to directly
employ 2D SR algorithms. 
Existing 3D SR algorithms, such as tempoGAN, can only be applied for relatively small up-scaling factors, whereas larger factors become \marie{utterly} expensive and difficult to train.
Larger factors directly \marie{imply} that the unknown function to be learned, i.e., the detail missing in the low-resolution (LR) input, contains \YouAdd{content with higher frequencies}. As such, \marie{the content} is more complex to represent \YouAdd{and it is more challenging to ensure its temporal coherence}. 
\YouAdd{A popular direction within the field of GANs is the 
progressive growing approach \cite{wang2018progressive}, which can achieve large up-scaling factors.}
However, the progressive growing of GANs has only been demonstrated for 2D content, as 3D volumetric data requires processing data and representing functions that have orders of magnitude more degrees of freedom.  
This is one of the key motivations for our approach:
leveraging multi-pass GANs to decrease the resource consumption for 
training \YouAdd{in order to} arrive at robust learning of 3D functions.

Simulating fluid flows at \YouAdd{high-resolution (HR)} is an inherently challenging process:
the \marie{number of} underlying numerical computations typically increase\MaxTodoNew{s} super-linearly when the 
discretization is refined. In addition to the spatial degrees of freedom, the temporal axis likewise needs to be refined to reduce numerical errors from time integration.
While methods like synthetic turbulence methods \cite{kim2008wavelet} typically rely on a finely resolved advection scheme, the deep learning\marie{-}based tempoGAN algorithm arrived at a frame-by-frame super-resolution scheme that circumvents the increase in computational complexity by working on independent frames of volumetric densities
via a spatio-temporal discriminator supervision.
Similar to tempoGAN, our multi-pass GAN also aims at fluid simulation SR, but targets two extra goals: reducing \MaxTodoNew{the unstable and intensive computations of the 3D training process} and increasing the range of possible up-scaling factors. To achieve those targets, multi-pass GAN adopts a divide-and-conquer strategy for 3D training. Instead of training with 3D volume data directly, multi-pass GAN divides the main task, \YouAdd{i.e.\marie{,} learning the 3D relationship between LR and HR, into two smaller sub-tasks by learning 2D relationships between LR and HR of two orthogonal planes.}
\marie{In a series of experiments,} we will demonstrate that this strategy strongly reduces the required computational resources and stabilizes the coupled non-linear optimization of GANs.

To summarize, the main contributions of our work are:
\vspace{-4pt}
\begin{itemize}
\item a novel multi-pass approach for robust network training with time sequences of 3D volume data,
\item extending progressive training approaches for multi-pass GANs, \marie{and}
\item combining progressive growing with temporal discriminators.
\end{itemize}
\vspace{-4pt}
In this way, we arrive at a first method for learning $8\times$ up-scaling of \marie{3D} fluid simulations
with deep neural networks.

\begin{figure*}
\begin{center}
\vspace{-10pt}
\includegraphics[width=0.9\linewidth]{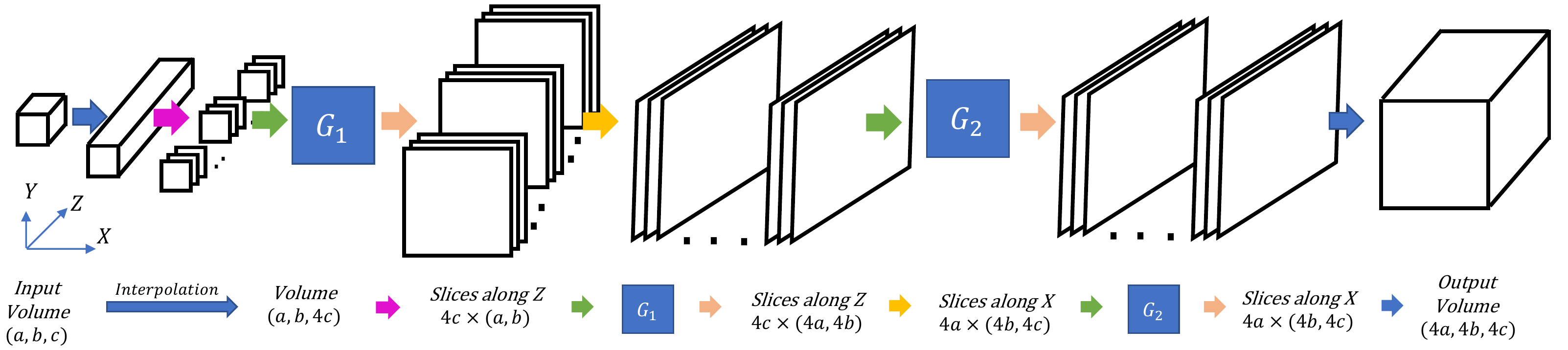}
\end{center}
   \vspace{-15pt}
   \caption{{\small The pipeline of the proposed multi-pass GAN.
   After an up-sampling along z, we process two orthogonal directions with two adversarially
   trained generator networks $G_1$ and $G_2$. The initial up-sampling ensures that all unknowns are processed
   evenly by the networks.}
   }
   \vspace{-8pt}
\label{fig:pipeline}
\end{figure*}

\vspace{-6pt}
\section{Related Work}
Large disparities between HR screens in household devices and LR data have driven \YouAdd{a large number of} advances in SR algorithms, which also find applications in other fields like surveillance \cite{zhang2010super} and medical image processing\cite{kennedy2006super}.
As deep learning\marie{-}based methods were shown to generate state-of-the-art results for SR problems, we focus on this class of algorithms in the following.
At first, convolutional neural networks (CNNs) were used in SRCNN \cite{dong2014learning}, and were shown to achieve better performance than traditional SR methods \cite{timofte2014a+}.
Theoretically, deeper CNNs can solve more arduous tasks, but
overly deep CNNs typically cause vanishing gradient problems. Here, CNNs with skip connections, such as residual networks \cite{he2016identity}, were applied
to alleviate \YouAdd{this problem}\cite{lim2017enhanced}. Instead of generating images directly, VDSR ~\cite{kim2016accurate} and ProSR~\cite{wang2018progressive}
\marie{apply} the CNNs \marie{to} learn the residual content, which largely 
\YouAdd{decreases the workload \marie{for} the networks and 
improves the quality of the results}. In our structure, we also applied residual learning to improve performance and reduce resource consumption.

Beyond architectures, choosing the right loss function is \marie{similarly} crucial.
Pixel-wise differences are \marie{the most basic} ones and are widely-used, \YouAdd{e.g., 
in DRRN~\cite{tai2017image}, LapSRN~\cite{lai2017deep} and SRDensseNet~\cite{tong2017image}.}
These methods achieve reasonable results and improve the performance in terms of the peak signal-to-noise ratio (PSNR) and the structural similarity index (SSIM). \YouAdd{However, minimizing vector norm differences typically blurs out small scale details and features, which leads to high perceptual errors.} 
With the help of a pre-trained VGG network, perceptual quality can be improved by minimizing differences of feature maps and styles between generated results and references~\cite{johnson2016perceptual}.
However, VGG is trained with \marie{large} \YouAdd{amounts of labeled} 2D image data, and, for fluid-related data-sets, no pre-trained networks exist. Alternatively, GANs~\cite{goodfellow2014generative} can improve perceptual quality significantly. In addition to the target model as generator, a discriminator is trained to distinguish generated results \marie{from} references, which guide\marie{s} the generator to output \marie{more} realistic results. Instead of minimizing pixel-wise differences, GANs aim to generate results that follow the underlying data distribution of the training data. As such, GANs outperform traditional loss functions for multimodal problems~\cite{ledig2017photo}. Original GANs, e.g., DC-GAN\cite{RadfordMC15}, adopt an unconditional structure, \YouAdd{i.e.\marie{,} generators only process randomly-initialized vectors as inputs.}
By adding a conditional input, conditional GANs (CGAN)~\cite{mirza2014conditional} learn the relationship between the condition and the target. Therefore, this input can be used to control the generated output.
By using conditional adversarial training, EnhanceNet~\cite{sajjadi2017enhancenet} and SRGAN~\cite{ledig2017photo} arrive at detailed results for natural images. tempoGAN \cite{xie2018tempogan} uses the velocity of the fluid simulation as a conditional input. This allows the network to learn the underlying physics and improves the result\marie{ing} quality.

For SR problems, larger up-scale factors 
typically lead to substantially harder learning tasks and potentially 
\YouAdd{more unstable training} \marie{processes} for recovering the details of the reference. Hence, most previous methods 
do not consider up-scaling factors larger than $2$ or $4$. However, ProSR~\cite{wang2018progressive} and progressive growing GANs~\cite{karras2017progressive} train the network in a staged manner, step-by-step. 
This makes it possible to generate HR results with up-scaling factors of $8$ and above. \YouAdd{Here, we also apply the progressive training pipeline to show that our method can be applied to complex 3D flow data.}

The works discussed above focus on single image SR and do not take temporal coherence into account, which, however, is crucial for sequential and animation\marie{-}related SR applications.
One solution to keep results temporally coherent is using a sequence as input data and generating an output sequence all at once~\cite{saito2017temporal, yu2017seqgan}. This improves the temporal relationships, but frames can only be generated sequentially, and an extension to 3D data would require processing 4D volumes.
For models that generate individual frames independently,
an alternative solution is to use additional temporal losses.
E.g., L2 losses were used to minimize differences between warped nearby frames \cite{ruder2016artistic, chen2017coherent}. However, this typically leads to sub-optimal perceptual quality. Beyond the L2 loss, \marie{specifically} designed loss term\marie{s} can be used to penalize discontinuities between adjacent frames~\cite{bhattacharjee2017temporal}. Instead of using explicit loss functions to restrict temporal relationships, temporal discriminators \cite{xie2018tempogan, chu2018temporally} can automatically supervise \nilsNew{w.r.t. temporal aspects}. \YouAdd{They also improve the perceptual quality of the generated sequences. In our work, a variant of such a temporal discriminator architecture is used.}

\YouAdd{To arrive at} realistic fluid simulations, methods typically aim for solving the Navier-Stokes (NS) equations and inviscid Euler equations as efficiently and precisely as possible. 
Based on the first stable single-phase fluid simulation algorithm~\cite{stam1999stable} for computer animation, a variety of extensions was proposed, e.g., more accurate advection schemes~\cite{kim2005flowfixer, selle2008unconditionally}, coupling between fluid and solids \cite{batty2007fast, teng2016eulerian}, and fluid detail synthesis based on turbulence models~\cite{kim2008wavelet, narain2008fast}.
In recent years, deep learning methods also attracted attention in the field of computer graphics, such as rendering \cite{bako2017kernel, chaitanya2017interactive}, volume illumination \cite{kallweit2017deep}\marie{,} and character control \cite{peng2017deeploco}. Since fluid \MaxTodoNew{simulations are} very time-consuming, CNNs are also applied to estimate parts of numerical calculations, \YouAdd{e.g., a CNN-based PDE solver~\cite{farimani2017deep, long2017pde, tompson2017accelerating} was proposed, as well as a fast SPH method using regression forests for velocity prediction~\cite{jeong2015data}}. CNNs can learn the relationships between control parameters and simulations of interactive liquids~\cite{prantl2017pre}. Neural networks are also used for splash generation~\cite{um2018liquid} and descriptor-encoding of pre-computed fluid patches for fluid synthesis~\cite{chu2017data}.
Generative networks were additionally trained to pre-compute solution spaces for smoke and liquid flows~\cite{kim2018deep}. 
In contrast to these methods, we aim for particularly large up-scaling models that could not be realized with the aforementioned methods.

\section{Method}\label{sec:method}

\YouAdd{Training neural network\marie{s} with a large number of weights, as \marie{it} is usually required
for 3D problems, is inherently expensive. 
\marie{In an adversarial setting training large networks is particularly challenging}, as stability is threatened
by the size of the \marie{volumetric} data. This in turn often leads to small mini-batch sizes and correspondingly unreliable gradients. 
In the following, we will explain how these difficulties of 3D GANs can be circumvented with our
multi-pass GAN approach.}

\vspace{-6pt} 
\subsection{Multi-Pass GAN}

\setlength{\columnsep}{4pt}%
\begin{wrapfigure}{hR}{0.60\linewidth} 
\centering
\vspace{-6pt} 
\includegraphics[width=0.99\linewidth]{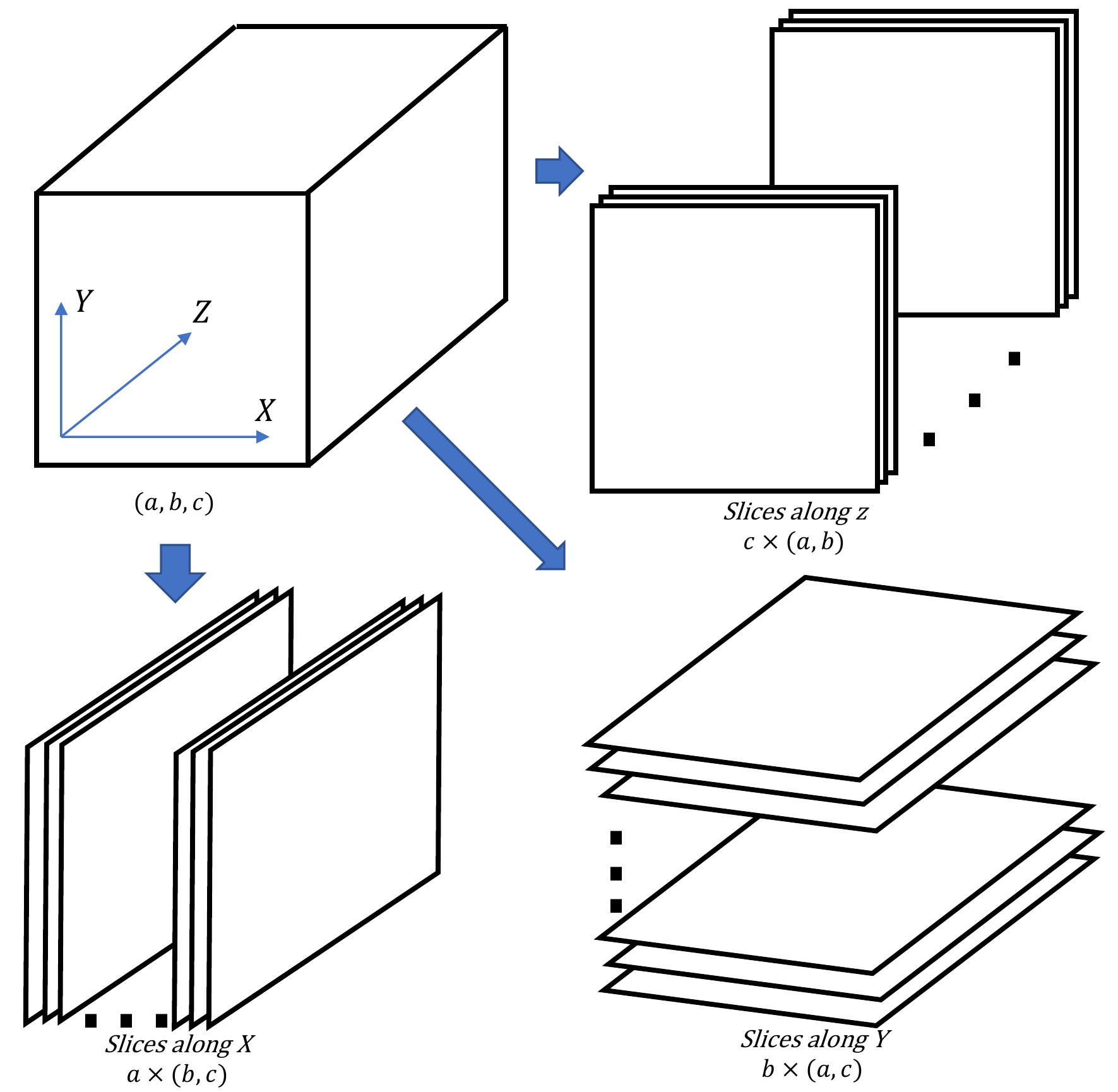}
\vspace{-24pt}
\caption{{\small Three \marie{alternatives} to decompose a 3D volume into stacked slices.}}
\vspace{-6pt}
\label{fig:slice}
\end{wrapfigure}
\setlength{\columnsep}{12pt}%

To reduce the dimensionality of the learning problem, we decompose the 3D data into a stack of 2D slices and train our networks on those slices using convolutions along two spatial dimensions. As shown in \myreffig{fig:slice}, there are three different ways to split a 3D volume into stacked slices. If we split the 3D volume along the $X$-axis, we can retrieve motion information in the $YZ$-plane for every $YZ$-slice, while splitting 
along the $Y$-axis yields information about the $XZ$-planes.

\newcommand{\myv}{1.0\linewidth}
\newcommand{\myw}{0.20}
\begin{figure}[b]
\begin{center}
\vspace{-6pt}
\begin{minipage}{\myw\linewidth} 
\vspace{0cm}
\includegraphics[trim={2px 2px 2px 2px}, clip, width = \myv]{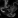} \\
\includegraphics[trim={2px 2px 2px 2px}, clip, width = \myv]{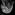} \\
\begin{overpic}[width=\myv,tics=10,clip,trim={3px 3px 3px 3px}]{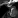}
 \put (4,4) {\textcolor{white}{$X$}}
 \end{overpic} 
\end{minipage}%
\begin{minipage}{\myw\linewidth}
\includegraphics[trim={8px 8px 8px 8px}, clip, width = \myv]{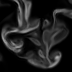} \\
\includegraphics[trim={8px 8px 8px 8px}, clip, width = \myv]{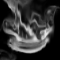}  \\
\begin{overpic}[width=\myv,tics=10,clip,trim={12px 12px 12px 12px}]{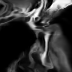}
 \put (4,4) {\textcolor{white}{$G_1$}}
 \end{overpic} 
\end{minipage}%
\begin{minipage}{\myw\linewidth}
\includegraphics[trim={8px 8px 8px 8px}, clip, width = \myv]{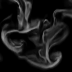} \\
\includegraphics[trim={8px 8px 8px 8px}, clip, width = \myv]{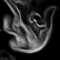}\\
\begin{overpic}[width=\myv,tics=10,clip,trim={12px 12px 12px 12px}]{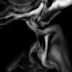}
 \put (4,4) {\textcolor{white}{$G_2$}}
 \end{overpic} 
\end{minipage}%
\begin{minipage}{\myw\linewidth}
\includegraphics[trim={8px 8px 8px 8px}, clip, width = \myv]{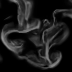} \\
\includegraphics[trim={8px 8px 8px 8px}, clip, width = \myv]{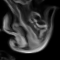} \\
\begin{overpic}[width=\myv,tics=10,clip,trim={12px 12px 12px 12px}]{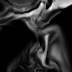}
 \put (4,4) {\textcolor{white}{$G_3$}}
 \end{overpic} 
\end{minipage}%
\begin{minipage}{\myw\linewidth}
\includegraphics[trim={8px 8px 8px 8px}, clip, width = \myv]{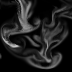} \\
\includegraphics[trim={8px 8px 8px 8px}, clip, width = \myv]{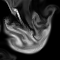}\\
\begin{overpic}[width=\myv,tics=10,clip,trim={12px 12px 12px 12px}]{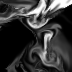}
 \put (4,4) {\textcolor{white}{$Y$}}
 \end{overpic} 
\end{minipage}
\end{center}
   \vspace{-12pt}
   \caption{{\small Comparison of the %low-resolution
   LR input density $X$, the output of the first network $G_1$, the second network $G_2$, and the third (optional) network $G_3$. The HR reference $Y$ is shown on the right. The first row depicts a slice of the $XY$-plane, the second \marie{row} one \marie{slice} of the $YZ$-plane\marie{,} and the third \marie{row} one \marie{slice} of the $XZ$-plane. While $G_2$ significantly improves the output, $G_3$ is largely redundant.} }
   \vspace{-6pt}
\label{fig:4xcomp1}
\end{figure}

In order to reconstruct the target 3D function when training with 2D slices, 
we perform the steps illustrated in \myreffig{fig:pipeline}.
First, we have to ensure that there is no bias in the inference, i.e., all spatial dimensions
are processed in the same manner.
Thus, for the LR volume $x$ with resolution ($a,b,c$), we linearly interpolate along the $Z$-axis to increase the resolution from ($a,b,c$) to ($a,b,4c$).
We then split the new volume along the $Z$-axis to obtain $4c$ slices of size ($a,b$). Here\marie{,} we train a 2D network $G_1$ to up-scale these ($a,b$) slices to the target size of ($4a,4b$). This first generator produces HR volumes, but as it only works in the $XY$-plane, the results are only temporally coherent in this plane,
and the model would still generate stripe-like artifacts in the $XZ$- or $YZ$-planes of the volumes. 
\marie{In order to allow the network to} evaluate motion information along the $Y-$ and  $Z$-dimensions, we use a second generator network $G_2$ to refine the details of \marie{the volume along} the $Y$- and $Z$-axes. Here\marie{,} we cut the volume along the $X$-axis to obtain $4a$ slices of size ($4b,4c$). $G_2$ is then trained to refine those slices and drive the distribution of the output closer to the target function, i.e.\marie{,} a\marie{n} HR simulation. Note that we have found it beneficial to not \marie{change the resolution with} $G_2$, but instead let it process the full resolution data.
In summary, combining the interpolation along $Z$, $G_1$ and $G_2$ jointly up-scale 3D volumes from ($a,b,c$) to ($4a,4b,4c$) and ensure that all voxels are consistently processed by \marie{the} two generator networks.

Theoretically, as soon as we have motion from two orthogonal directions, we \marie{are able to infer} the full 3D motion. 
Thus\marie{,} we train $G_1$ and $G_2$ with $XY$- and $YZ$-slices, respectively, such that they -- in combination -- \marie{obtain} full motion information.
To verify that this is sufficient, we also employed a third network $G_3$ to refine the $XZ$-plane with an additional pass \marie{through} training \marie{on} $XZ$-slices.
Results from the first, second\marie{,} and this optional third network can be found in \myreffig{fig:4xcomp1}, where each row illustrates examples of a different slicing direction. In the first row, for example, the volume is cut along the $Z$-axis, which $G_1$ operates on. When comparing the second column to the third one, it is notable that there is not much improvement for the $XY$-plane. However, looking at the second row which is a sample of the $YZ$-plane that $G_2$ focuses on, it is clear that the artifacts generated by the linear interpolation are removed and the spatial coherence along the $Z$-axis \marie{is} substantially improve\marie{d}. Even the slices in the $XZ$-plane \marie{are enhanced}, \marie{as shown in} the third row \marie{of} \myreffig{fig:4xcomp1}. 
The fourth column in the \myreffig{fig:4xcomp1} shows that $G_3$ leads to minimal changes along any of the axes, which matches \marie{our assumption} that this additional direction of slicing is redundant.
\marie{Therefore,} we focus on two generator networks \marie{for our final structure}, $G_1, G_2$, while $G_3$ is not applied.
\\

\YouAdd{\section{Multi-pass GAN with Temporal Coherence}\label{sec:tempoGAN}
In general, the super-resolution problem is a multimodal task and here our goal is to additionally obtain physically plausible results with a high perceptual quality. Thus, we employ adversarial training to train both $G_1$ and $G_2$. 
The full optimization problem for regular GANs can be formulated as \cite{goodfellow2014generative}:
\resizeEq{
\min_{G}\max_{D} V(D, G))= & \mathbb{E}_{y\sim p_{\text{y}}(y)}[\log D(y)] +\mathbb{E}_{x\sim p_{\text{x}}(x)}[\log (1-D(G(x)))] \\
= & \myavg_m [\log D(y_m)] + \myavg_n [\log (1-D(G(x_n)))],
}{eq:GANloss}{0.9}
where $m, n$ denote the number of reference samples $y$ and drawn inputs $x$\marie{,} respectively. $y\sim p_{\text{y}}(y)$ \marie{states} that reference data is sampled from the probability distribution $p_{\text{y}}$.

\marie{Our goal is similar to the one from} the tempoGAN network~\cite{xie2018tempogan}. \marie{To} illustrate the advantages of our approach, we will first employ \marie{our network} in a tempoGAN pipeline for $4\times$ fluid SR in this section
before targeting large up-scale factors in conjunction with a progressive growing scheme in \myrefsec{sec:growinggan}.  
\subsection{Network Structure and Loss Function}\label{sec:sttempoGAN}}

\setlength{\columnsep}{4pt}%
\begin{wrapfigure}{hR}{0.68\linewidth} 
\centering
\vspace{-6pt} 
\includegraphics[width=0.99\linewidth]{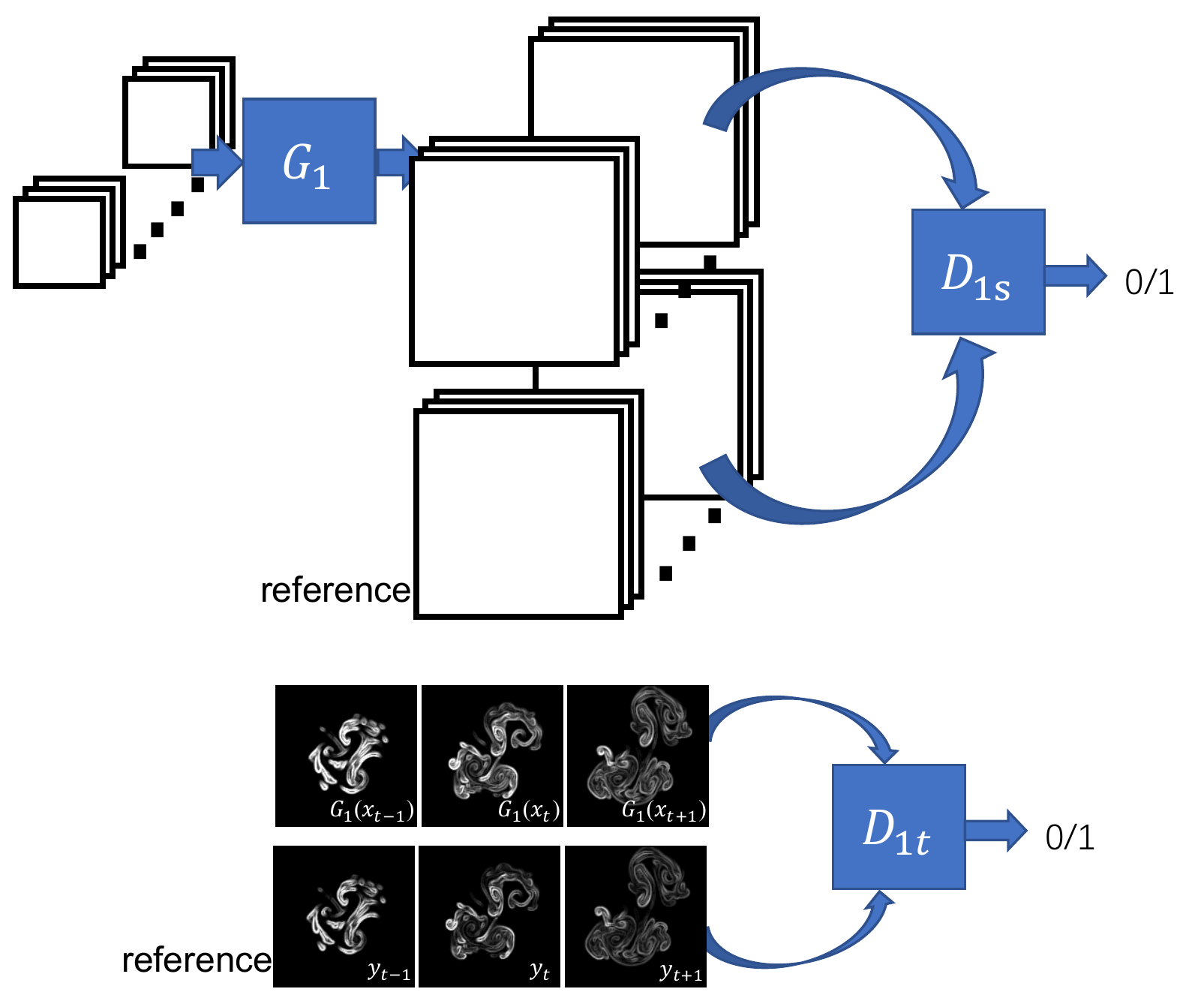}
\vspace{-24pt}
\caption{{Spatial and temporal discriminators $D_{1s}$ and $D_{1t}$ supervise $G_1$ on the $XY$-plane in a multi-pass tempoGAN.}}
\vspace{-6pt}
\label{fig:G&D}
\end{wrapfigure}
\setlength{\columnsep}{12pt}

\begin{figure*}
    \vspace{-6pt}
    \centering
    \newcommand{\myvv}{0.18\textwidth}
    \begin{tabular}{@{}c@{}c@{}c@{}c@{}c@{}}
    
\begin{minipage}{0.24\linewidth} 
\begin{overpic}[width=\myv,tics=10,clip,trim={  700px, 200px, 220px, 180px}]{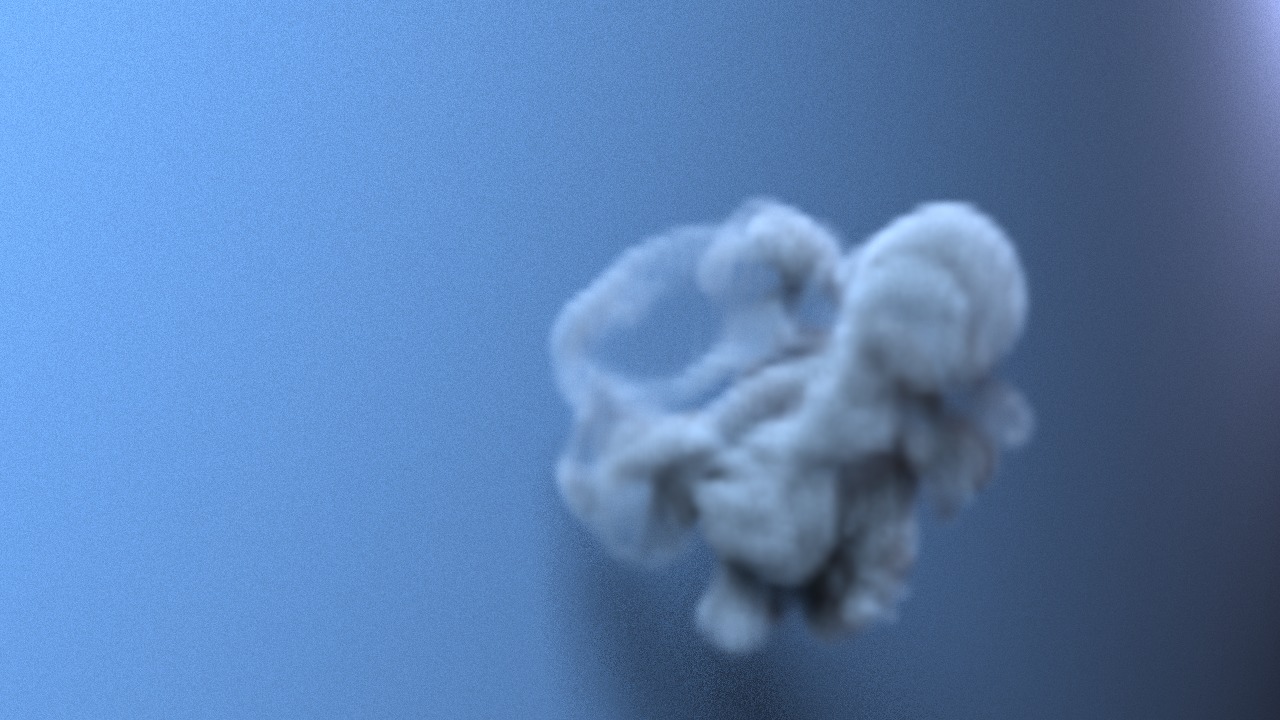}
 \put (4,4) {\textcolor{white}{$x$}}
 \end{overpic} 
\end{minipage}%
\begin{minipage}{0.24\linewidth} 
\begin{overpic}[width=\myv,tics=10,clip,trim={  700px, 200px, 220px, 180px}]{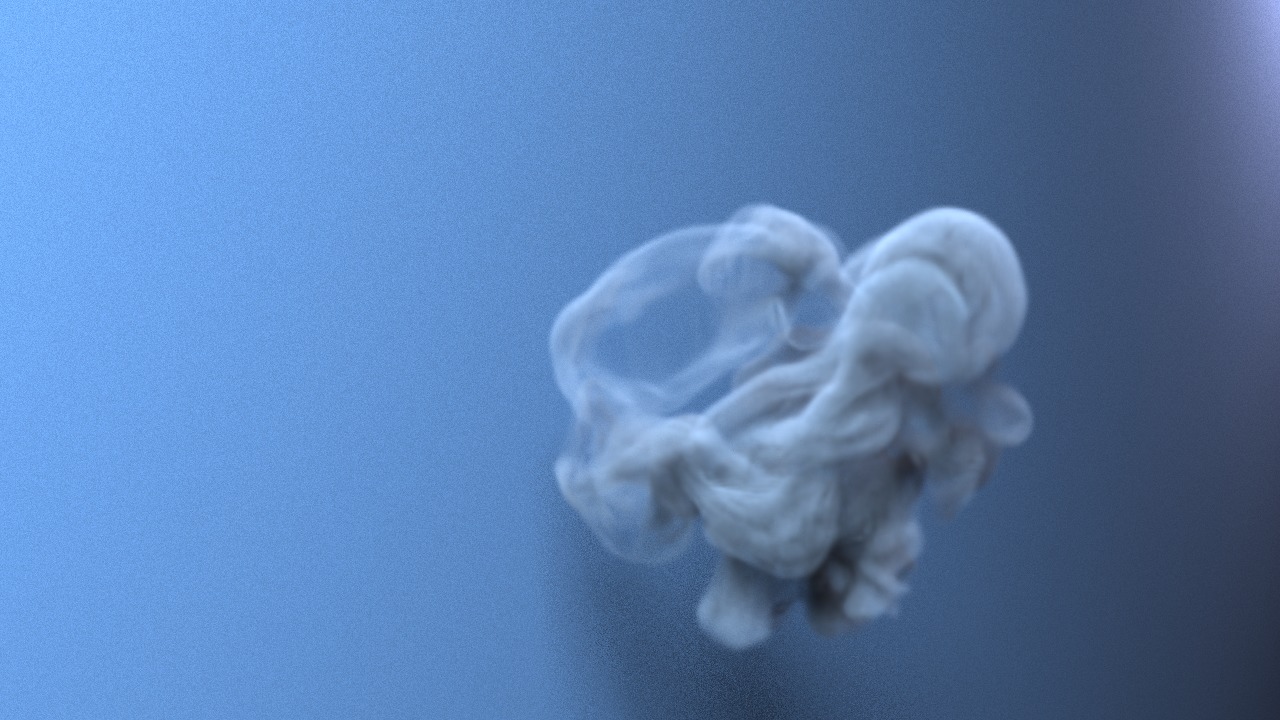}
 \put (4,4) {\textcolor{white}{$G_2$}}
 \end{overpic} 
\end{minipage}%
\begin{minipage}{0.24\linewidth} 
\begin{overpic}[width=\myv,tics=10,clip,trim={  700px, 200px, 220px, 180px}]{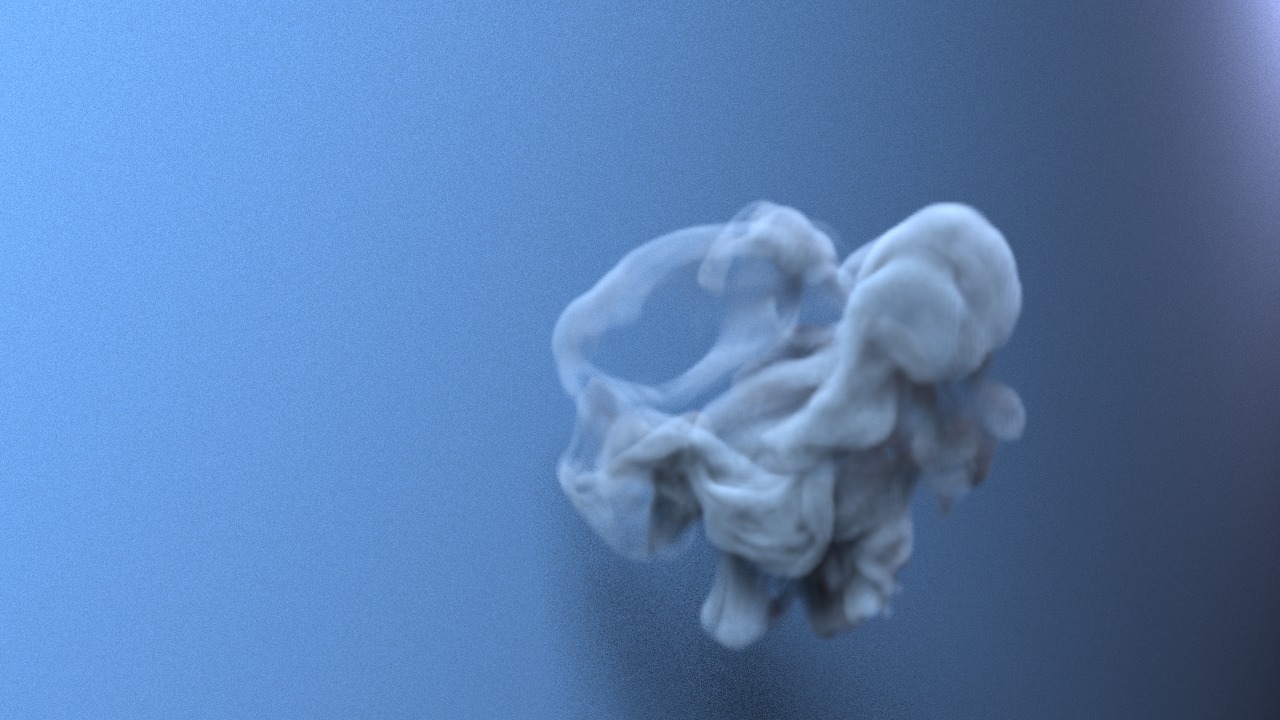}
 \put (4,4) {\textcolor{white}{\cite{xie2018tempogan}}}
 \end{overpic} 
\end{minipage}%
\begin{minipage}{0.24\linewidth} 
\begin{overpic}[width=\myv,tics=10,clip,trim={  700px, 200px, 220px, 180px}]{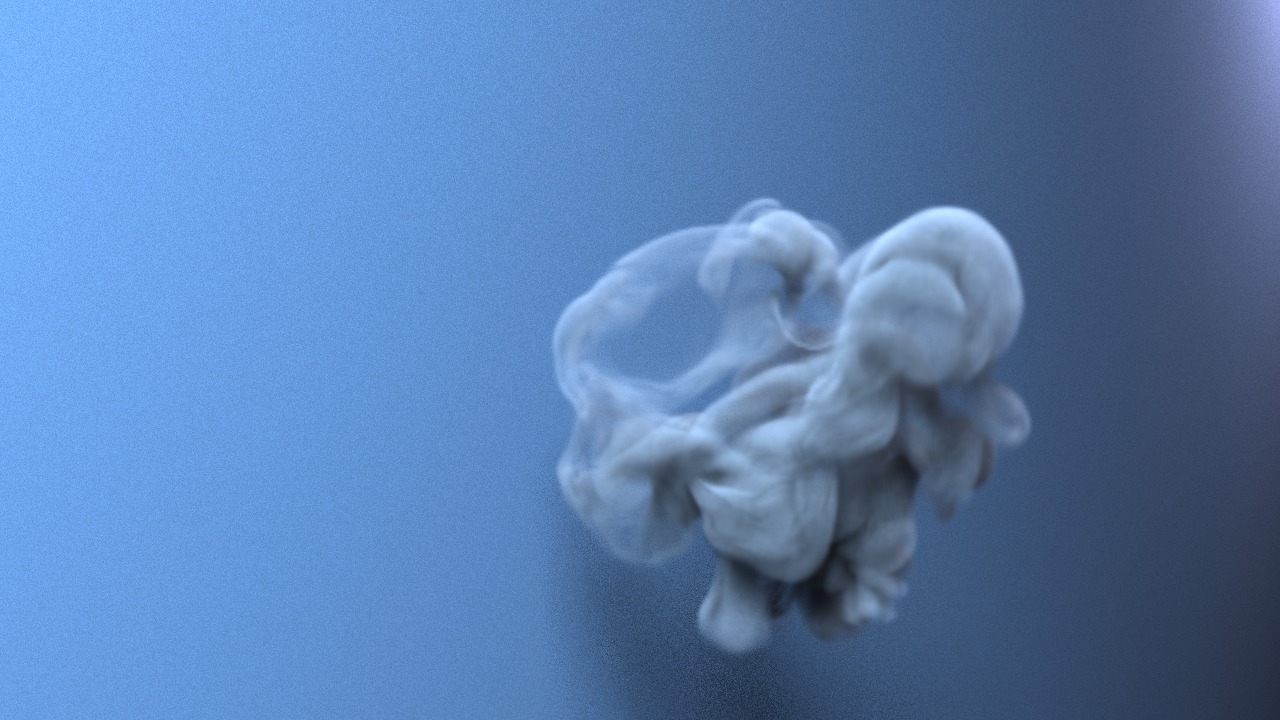}
 \put (4,4) {\textcolor{white}{$y$}}
 \end{overpic} 
\end{minipage}%

    \end{tabular}
    \vspace{-6pt}
    \caption{\small {Comparison of LR input, output of $G_2$, regular 3D tempoGAN and HR reference. The latter three have the same resolution. }}
    \label{fig:rendered4xcomp3}
\end{figure*}
\begin{figure}[b]
    \centering
\begin{overpic}[trim={900px, 300px, 280px, 320px}, clip, width = 0.11\textwidth]{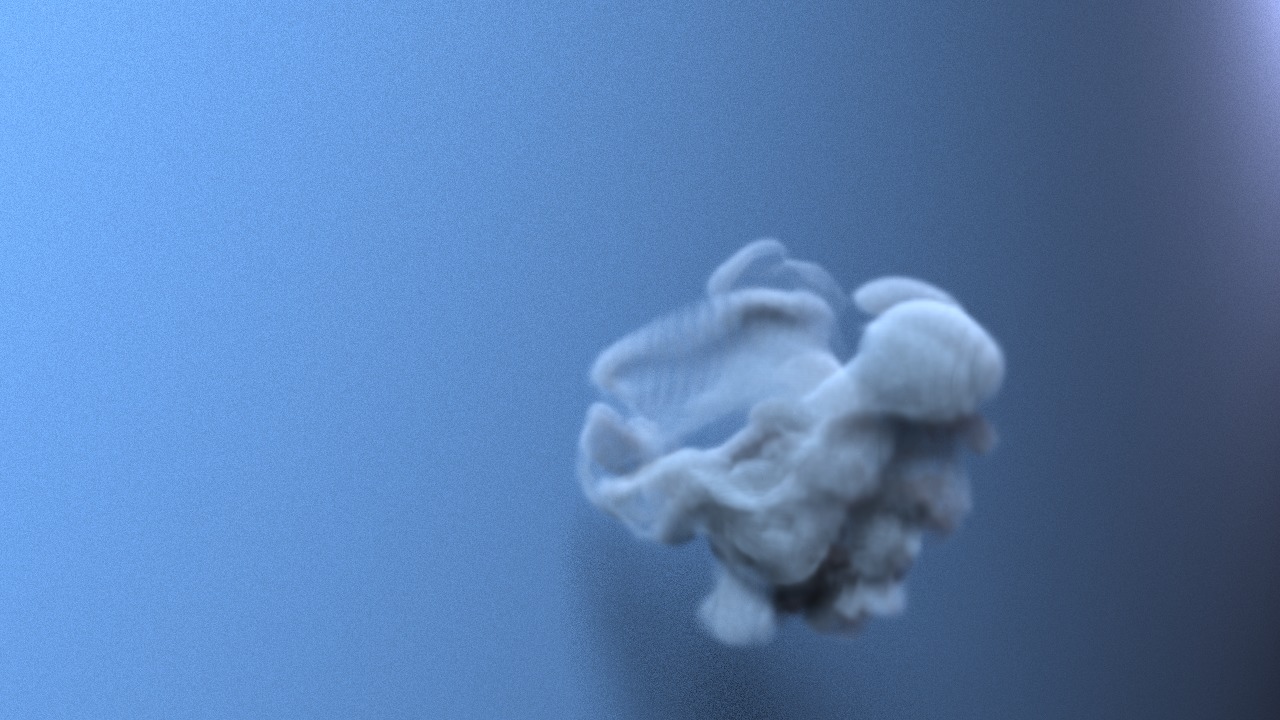} \put (4,4) {\small\textcolor{white}{a) $G_1$}}
 \end{overpic} 
\begin{overpic}[trim={900px, 300px, 280px, 320px}, clip, width = 0.11\textwidth]{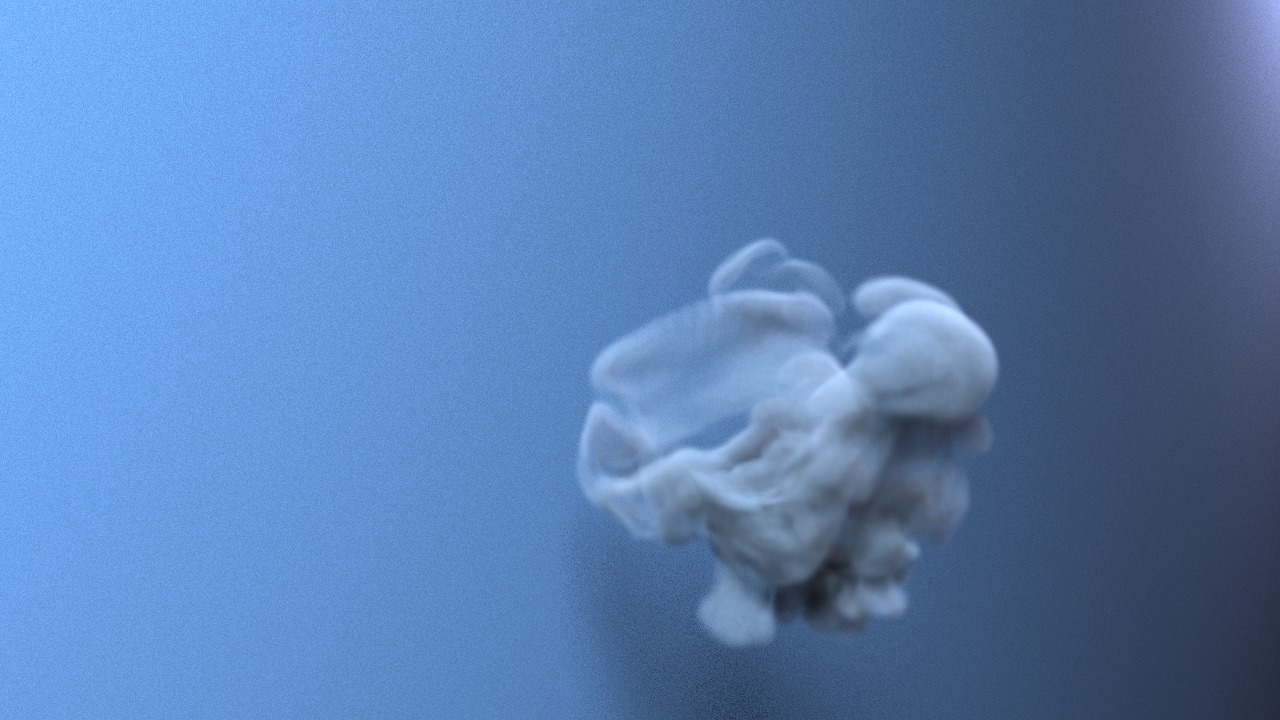} 
 \put (4,4) {\small\textcolor{white}{a) $G_2$}}
 \end{overpic} %\quad
\begin{overpic}[trim={220px, 100px, 120px, 240px}, clip, width = 0.11\textwidth]{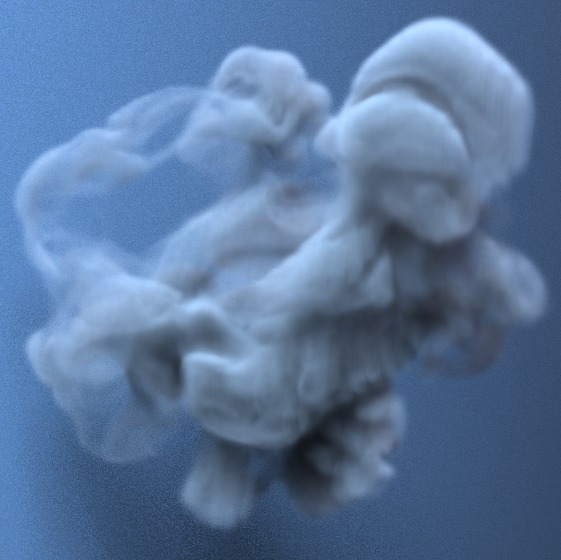} \put (4,4) {\small\textcolor{white}{b) $G_1$}}
 \end{overpic} 
\begin{overpic}[trim={220px, 100px, 120px, 240px}, clip, width = 0.11\textwidth]{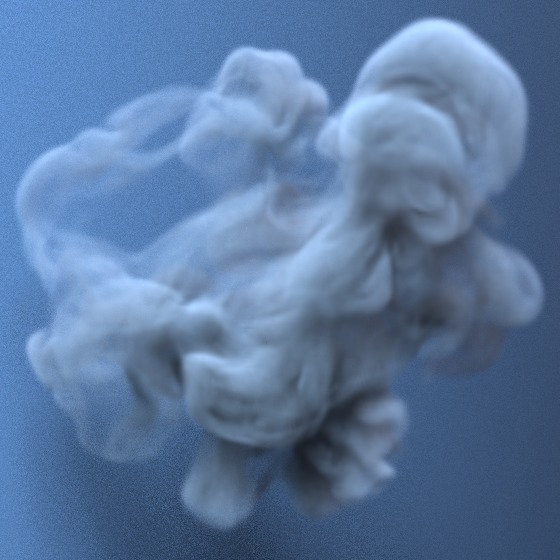}
 \put (4,4) {\small\textcolor{white}{b) $G_2$}}
 \end{overpic} 
    \vspace{-6pt}
   \caption{{\small Example volumes generated with a multi-pass tempoGAN, shown after the first and second pass. The application of $G_2$ removes the stripe-like artifacts from $G_1$ in a) and adds new details along the yet unseen axis in b). \MaxTodoNew{All examples were extracted from former frames of the simulation shown in \myreffig{fig:rendered4xcomp3} to \marie{emphasize} the clear differences between the passes}.}} 
\label{fig:rendered4xcomp2}
\end{figure} 
Unlike \marie{regular} GANs, 
which only contain one discriminator, tem\-po\-GAN uses two discriminators\marie{:} one spatial discriminator $D_s$ to constrain spatial details, and one temporal discriminator $D_t$ to keep the sequence temporally coherent.
For $D_t$, three warped adjacent frames' densities are used as inputs to 
\marie{enforce} learn\marie{ing} temporal evolution in the discriminator. The tempoGAN generator furthermore requires velocity information as input in order to distinguish different amounts of turbulent detail to be generated. \MaxTodoNew{Similar \marie{to} tempoGAN, our generator model consists of 4 residual blocks and takes the 
LR density and velocity fields as inputs, which, assuming that the fields are of size $16^2$, leads to an input of size $16\times16\times4$, with one density and three velocity channels.}

For a multi-pass version of tempoGAN, there are two generators\marie{:} $G_{1}$ and $G_{2}$. Every generator is paired with a spatial and a temporal discriminator, i.e., $D_{1s}$, $D_{1t}$ and $D_{2s}$, $D_{2t}$, respectively.
The training process of $G_1$ is shown in \myreffig{fig:G&D}. For $G_2$, a similar procedure is applied, \marie{with} the only difference being the up-scale operation in the network and the input data which consists of the output of the first network and LR velocities.

\YouAdd{During \MaxTodoNew{the} training of $G_1$ and $G_2$,} we use an 
additional L1 loss and feature space loss terms to stabilize the training.
The resulting loss functions for training $G_{1}$, $D_{1s}$, $D_{1t}$ are then given by:
\resizeEq{
\mathcal{L}_{G_1}(D_1s,D_1t,G_1)=& -\myavg_n [\log D_1s(x, G_1(x)) ]
- \myavg_n [\log D_1t\Big(\widetilde{G_1}_{\mathcal{A}}\Big(\widetilde{X}\Big)\Big)] \\
& + \myavg _{n,j} \lambda_{f}^{j} \left \| F^{j}(G_1(x))-F^{j}({y}) \right \|_{2}^{2} + \lambda _{L_1}  \myavg_n \left \| G_1(x)-{y}\right \|_{1}
}{eq:4xlossfunction1}{0.92}
\vspace{-6pt}
\resizeEq{
\mathcal{L}_{D_1t}(D_1t,G_1)= & -\myavg_m [\log D_1t(\widetilde{Y}_{\mathcal{A}})] - \myavg_n [\log \Big(1-D_1t\Big(\widetilde{G_1}_{\mathcal{A}}\Big(\widetilde{X}\Big)\Big)\Big)]\\
\mathcal{L}_{D_1s}(D_1s,G_1)=& -\myavg_m [\log D_1s(x, y)] - \myavg_n [\log (1-D_1s(x, G_1(x)))]\marie{,}
}{eq:4xlossfunction}{0.92}
where $\mathcal{A}$ advects a given frame with the current velocity, \marie{s.t.} $y^{t} = \mathcal{A}( y^{t-1}, v_y^{t-1} )$. $\widetilde{G}_{\mathcal{A}}(${\footnotesize{$\widetilde{X}$}}$)$ denotes three advected, consecutive generated frames: $\widetilde{G}_{\mathcal{A}}(${\footnotesize{$\widetilde{X}$}}$) = \{ \mathcal{A}( G(x^{t-1}), v_{x}^{t-1})$, $G(x^{t})$, 
$\mathcal{A}( G(x^{t+1}), -v_{x}^{t+1}) \}$, while $\widetilde{Y}_{\mathcal{A}}$ denotes three ground truth frames:
$\widetilde{Y}_{\mathcal{A}} = \{ \mathcal{A}( y^{t-1}, v_x^{t-1} )$, $y^{t}$, $\mathcal{A}( y^{t+1}, -v_x^{t+1} )\}$. $j$ is a layer in our discriminator network, and $F^{j}$ denotes the activations of the corresponding layer.  $G_{2}$, $D_{2s}$, and $D_{2t}$ are trained analogously.
Note that the fluid is only advected in the $XY$-plane and in the $YZ$-plane \marie{for $G_1$} and $G_2$\marie{, respectively}.

Exemplary outputs of $G_1$ and $G_2$ from a multi-pass tempoGAN are shown in \myreffig{fig:rendered4xcomp2}. 
While the output of $G_1$ still contains visible and undesirable interpolation artifacts, they are successfully removed by $G_2$.
\myreffig{fig:rendered4xcomp3} shows a comparison \marie{of} the original tempoGAN model and our multi-pass version
at the same target resolution. Our approach yields a comparable amount of detail and visual quality, while being the result of a much simpler and faster learning problem. We will evaluate this aspect of our approach in more detail below.
\MaxTodoNew{\\ \hfill } % refactoring...
\YouAdd{\section{Multi-pass GAN with Growing}\label{sec:growinggan}}

\begin{figure}[b]
\begin{center}
\vspace{-6pt}
\includegraphics[width=0.8\linewidth]{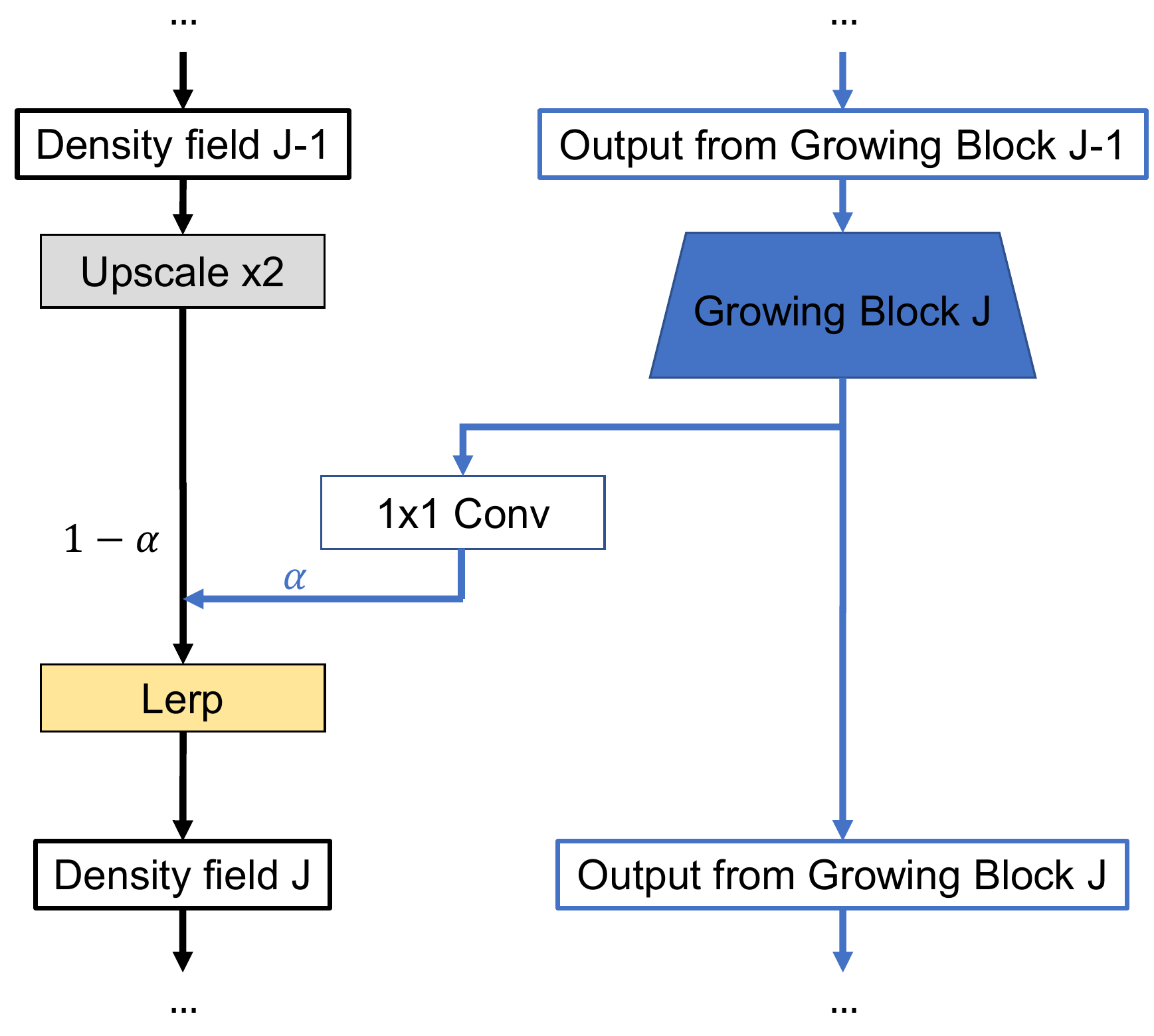}
\end{center}
   \vspace{-6pt}
   \caption{The structure of the progressive Growing GAN\marie{, where} the parameter $\alpha$ is used to interpolate between the up-scaled image from the former growing block and the current one. 
   For $\alpha=1$\marie{,} the former up-sampled density is \MaxTodoNew{ignored} completely.}
\label{fig:progan_structure}
\end{figure}

\begin{figure*}
    \centering
    \vspace{-3pt}
    \newcommand{\myvv}{0.18\textwidth}
    \begin{tabular}{@{}c@{}c@{}c@{}c@{}c@{}}
    
\begin{minipage}{0.24\linewidth} 
\begin{overpic}[width=\myv,tics=10,clip,trim={650px, 440px, 330px, 70px}]{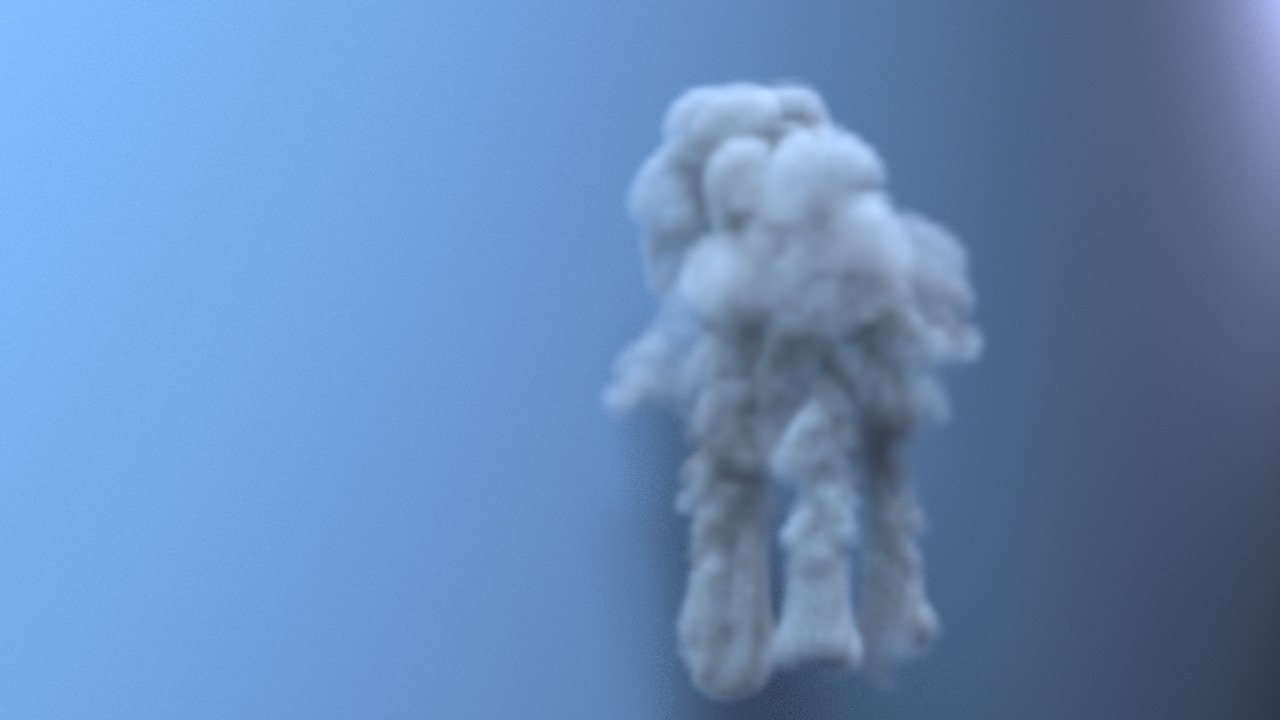}
 \put (4,4) {\textcolor{white}{$X$}}
 \end{overpic} 
\end{minipage}%
\begin{minipage}{0.24\linewidth} 
\begin{overpic}[width=\myv,tics=10,clip,trim={   650px, 440px, 330px, 70px}]{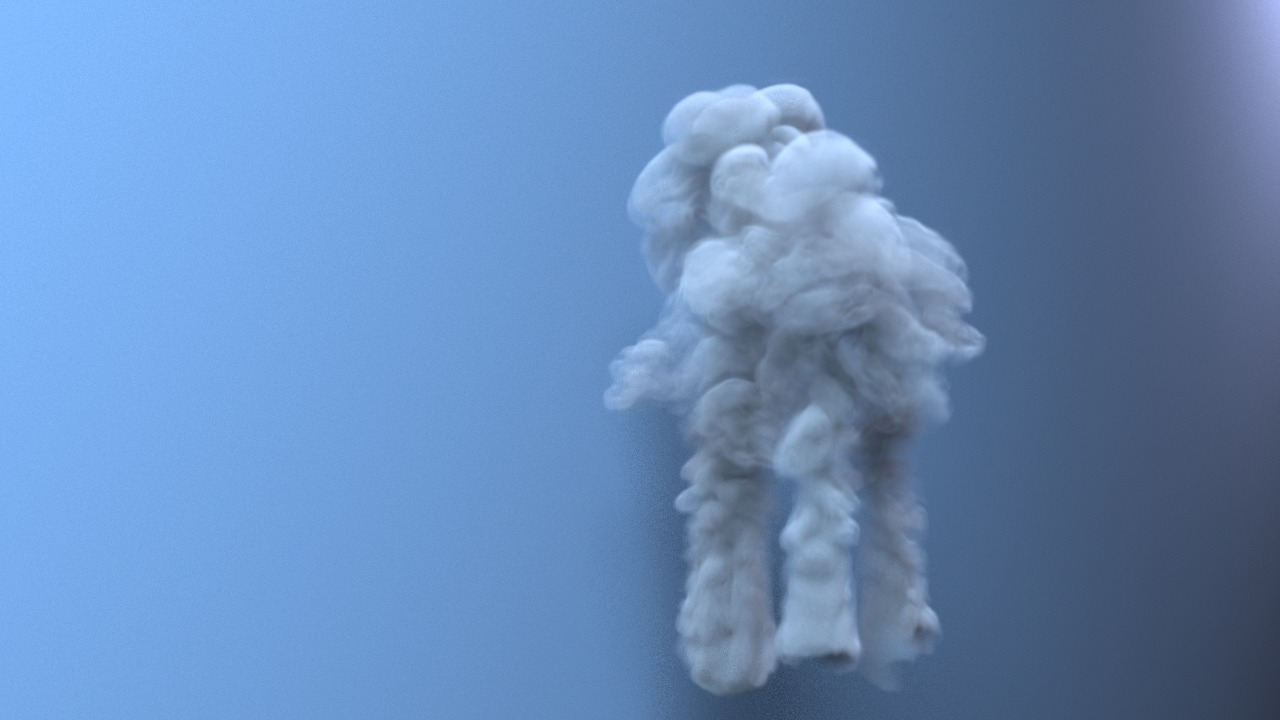}
 \put (4,4) {\textcolor{white}{$G_{1,8\times}$}}
 \end{overpic} 
\end{minipage}%
\begin{minipage}{0.24\linewidth} 
\begin{overpic}[width=\myv,tics=10,clip,trim={   650px, 440px, 330px, 70px}]{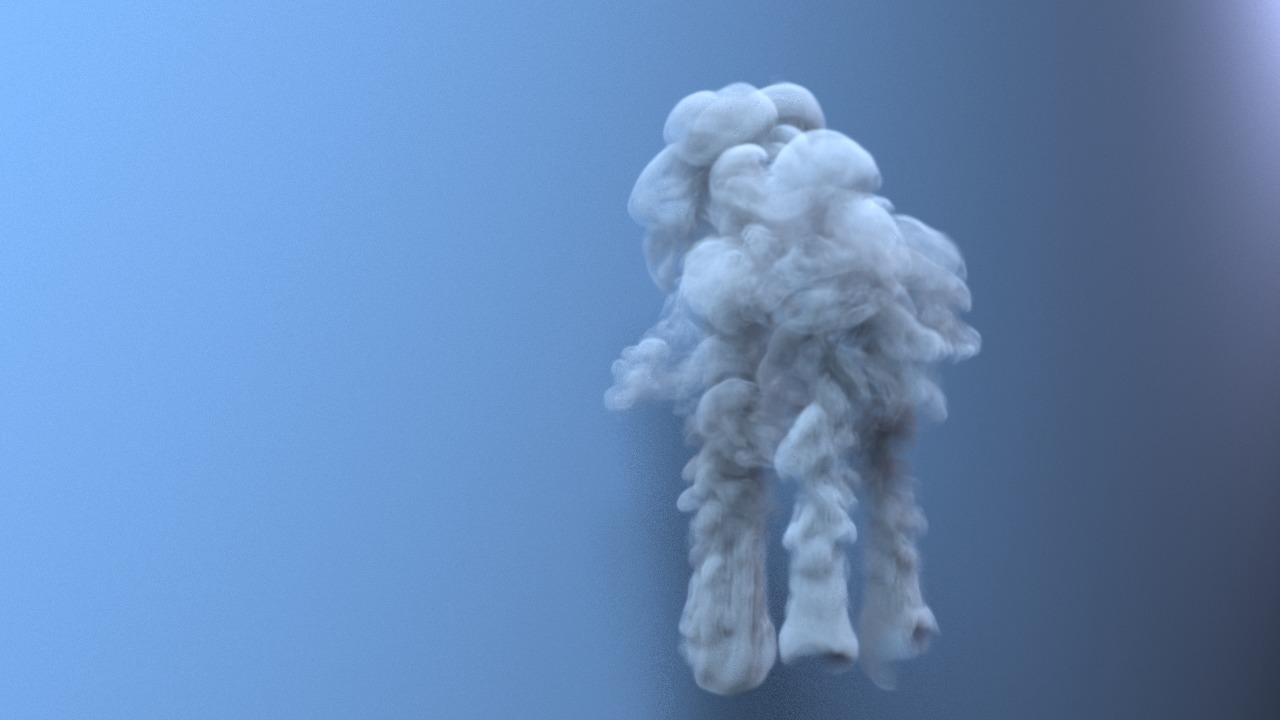}
 \put (4,4) {\textcolor{white}{$G_{2,8\times}$}}
 \end{overpic} 
\end{minipage}%
\begin{minipage}{0.24\linewidth} 
\begin{overpic}[width=\myv,tics=10,clip,trim={   650px, 440px, 330px, 70px}]{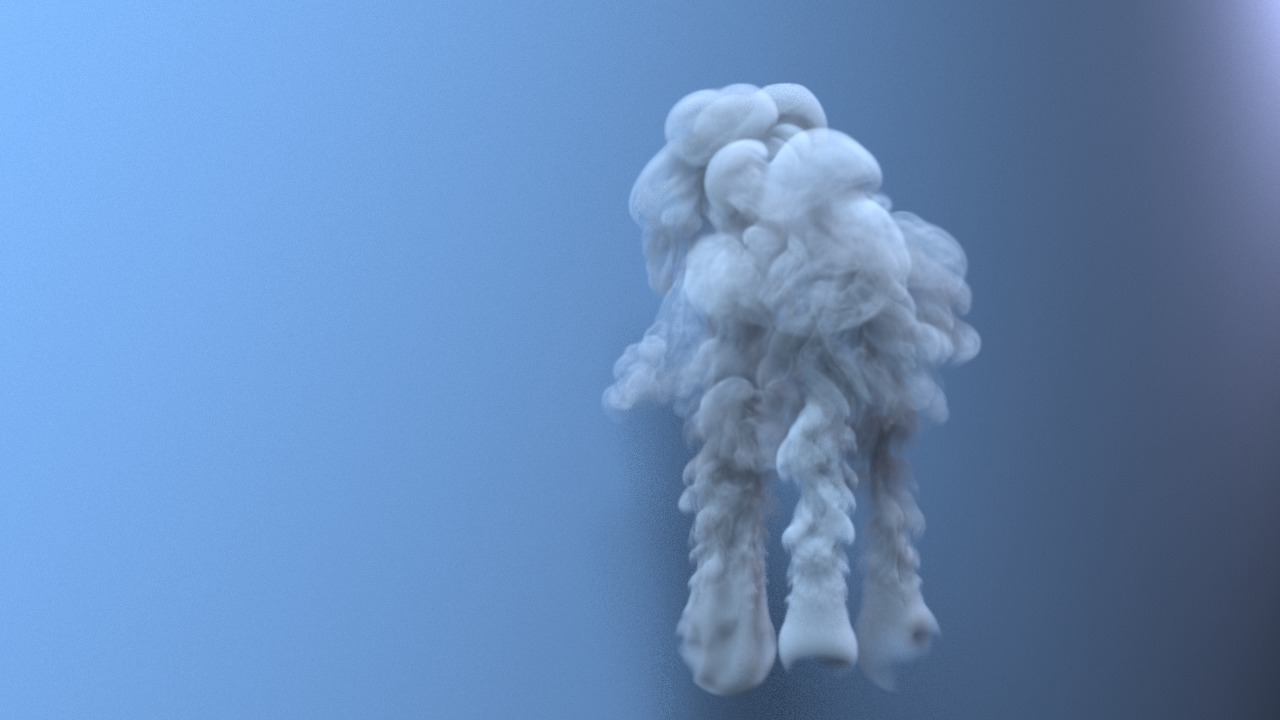}
 \put (4,4) {\textcolor{white}{$Y$}}
 \end{overpic} 
\end{minipage}%

    \end{tabular}
    \vspace{-3pt}
    \caption{\small {Comparison of LR input, output of $G_{1,8\times}$\marie{, output of} $G_{2,8\times}$\marie{,} and HR reference. The second generator $G_{2,8\times}$ is able to reconstruct missing smoke, sharpen smaller structures\marie{,} and add novel details to the volume.}}
    \label{fig:rendered8xcomp3}
\end{figure*}

\marie{Due to the coupled non-linear optimization involving} multiple networks\marie{,} GANs are particularly hard to train.
As this \marie{challenge} grows with larger numbers of weights in the networks, 
techniques such as curriculum learning \cite{wang2018progressive} and 
progressive growing \cite{karras2017progressive} were proposed to alleviate these inherent difficulties.
In the following, we will outline both techniques briefly before explaining how to adopt them
for our multi-pass GAN in order to increase the upscaling factor to $8$.

Curriculum learning describes the process of increasing the difficulty of a training target over time. 
\marie{In our case}, we first train the generator and discriminators on pre-computed density references that are twice as large as the input. %and which were generated beforehand. 
After training for a number of iterations, typically $120$k, 
we double the up-scaling factor until the final goal of $8$ is reached. 
The progressive growing approach goes hand in hand with curriculum learning: after increasing the difficulty of training, \marie{additional} layers \marie{are smoothly faded in} for the generator and discriminator networks. This ensures that the gradients from the added stages don't \marie{impair} the existing learning progress. As shown in \myreffig{fig:progan_structure}, the blending process is controlled by the parameter $\alpha\in[0,1]$, which is used to linearly interpolate between two current up-scaled density fields, \marie{e.g.}, combining generated outputs of scale $2$ and $4$. 
The procedure is similar \marie{regarding the discriminators}. However, we use average pooling layers to down-scale the density field between the stages instead of up-scaling it. 
\MaxTodo{This growing \marie{technique} is applied to the generator network $G_{1,8\times}$, the spatial discriminator $D_{1s,8\times}$\marie{,} as well as the temporal discriminator $D_{1t,8\times}$.}
While fading in new layers, we increase $\alpha$ from $0$ to $1$ over the course of $120$k iterations. 
The networks are then trained for another $120$k iterations on the current SR factor. 
\MaxTodo{Since the task for the second generator is to purely refine the volume along the axis that was \marie{invisible to} $G_{1,8\times}$\marie{,} and because the only training target is the HR reference, we disable progressive growing and curriculum learning for $G_{2,8\times}$, $D_{2s,8\times}$\marie{,} and $D_{2t,8\times}$.}\\ 

\subsection{Network Architecture}

The resulting generator network $G_{1,8\times}$ consists of a sequence of residual blocks (ResBlocks).
Overall, for a total SR factor of $8$, \marie{eight} ResBlocks are used which \marie{are} divided into \marie{four} growing blocks. Each growing block up-scales the previous layer by a factor of $2$, except for the first one \marie{which} maintains the input resolution.
For each growing block, a $1\times1$ convolution is trained in addition to translate the intermediate outputs into \marie{the} density field if the additional down-stream layers are not yet active as shown in \myreffig{fig:progan_structure}.
\nilsNew{For $G_{1,8\times}$\marie{,} \marie{in order to save computations,} we concentrate most of the parameters in the earlier stages of the network which have smaller spatial resolutions.}
We use a Wasserstein GAN loss with gradient penalty (WGAN) and a weak L1-loss in combination with the proposed equalized learning rate and pixelwise normalization (PN)~\cite{karras2017progressive} instead of batch normalization to keep the magnitudes in the adversarial networks under control. Thus, each ResBlock 
consists of the sequence of Conv3x3-ReLU-PN-Conv3x3-ReLU-PN.

In addition, inspired by VDSR ~\cite{kim2016accurate} and ProSR~\cite{wang2018progressive}, \MaxTodo{both our generators output} residual details rather than the whole density field to improve the quality of results and to decrease computing resources. Additionally, a bi-cubic up-scaled LR density field \marie{is} added to the output of $G_{1,8\times}$ and \MaxTodo{$G_{2,8\times}$. The architecture of $G_{1,8\times}$ leads to a receptive field of approximately $7$ LR pixels in both dimensions.
For $G_{2,8\times}$, $D_{2s,8\times}$\marie{,} and $D_{2t,8\times}$\marie{,} we increase the kernel size of the convolutional layers to $5$ in order to increase the receptive range to $4$ LR cells in each direction, resulting in a receptive field of $16$ LR cells. Furthermore, we decrease the amount of feature maps to keep the number of parameters similar to $G_{1,8\times}$, $D_{1s,8\times}$\marie{,} and $D_{1t,8\times}$. }

Beyond our multi-pass approach, the resulting network differs from existing growing approaches in \marie{the sense} that it targets functions of much larger dimensionality (3D plus time) via a temporal discriminator, while it also differs from the tempoGAN network in several ways: we employ a residual architecture with Wasserstein GAN loss \marie{and due to the progressive growing,} the networks are inherently different. 
The exact network structure and all training parameters can be found in Appendix~\ref{appendix:A} and ~\ref{appendix:B}. 

\nilsNew{\myreffig{fig:rendered8xcomp3} shows that the combination of our method and the growing approach can yield highly detailed and coherent buoyant smoke. $G_{2,8\times}$ is able to add additional details to the volume, reconstruct the desired HR shape, and generally sharpen the output along the axis that was unseen by $G_{1,8\times}$.}

\subsection{Loss Function}

\begin{figure}[b]
\begin{center}
\begin{overpic}[width=0.32\linewidth]{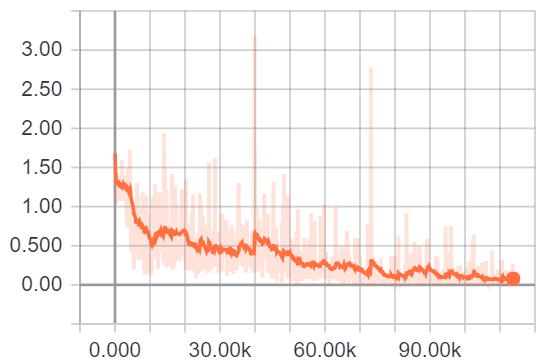}
 \put (5,2) {\footnotesize \textcolor{black}{a)}}
 \end{overpic} ~
 \begin{overpic}[width=0.32\linewidth]{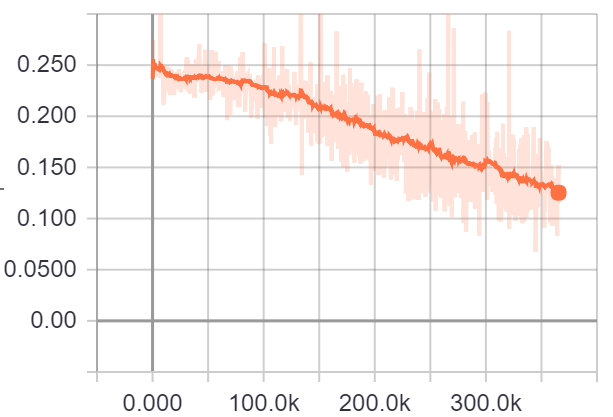}
 \put (5,2) {\footnotesize \textcolor{black}{b)}}
 \end{overpic} ~
 \begin{overpic}[width=0.32\linewidth]{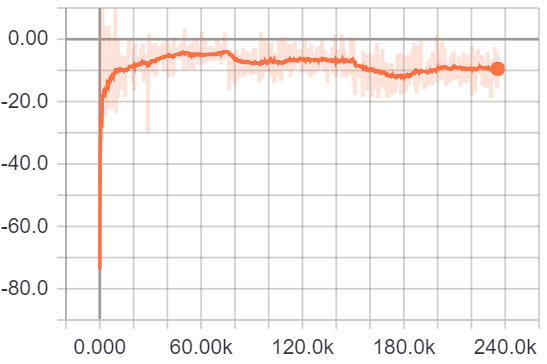}
 \put (3,2) {\footnotesize \textcolor{black}{c)}}
 \end{overpic} 
 \end{center}
 \vspace{-10pt}
 \caption{{\small Comparison of the spatial discriminator training loss when using a) the full TempoGAN loss, b) the LSGAN loss\marie{,} or c) the WGAN loss with gradient penalty. New layers were blended in at 60k iterations for b), c) and at 50k iterations for a).}}
\label{fig:loss_comparison}
\end{figure}

The resulting Wasserstein GAN\marie{-}based loss function for our multi-pass, multi-discriminator network 
is then given by:
\resizeEq{    
\mathcal{L}_{D_{1t,8\times}} (D_{1t,8\times},G_{1,8\times}) =&- \myavg_n[D_{1t,8\times}(\widetilde{Y}_{\mathcal{A}})]+\myavg_n[D_{1t,8\times}\Big(\widetilde{G_{1,8\times}}_{\mathcal{A}}\Big(\widetilde{X}\Big)\Big)]\\&+\lambda_{W} \myavg_n[(\vert\vert\nabla_{\hat{Y}_{\mathcal{A}}}D(\hat{Y}_{\mathcal{A}})\vert\vert_2-1)^2] \nonumber
}{eq:8xGrowingLoss1}{0.92}
\vspace{-6pt}
\resizeEq{   
\mathcal{L}_{D_{1s,8\times}}(D_{1s,8\times},G_{1,8\times})=&-\myavg_m[D_{1s,8\times}(x,y)]+\myavg_m[D_{1s,8\times}(x,{G_{1,8\times}}(x))]\\&+\lambda_{W}\myavg_m[(\vert\vert\nabla_{\hat{y}}D(\hat{y})\vert\vert_2-1)^2] 
}{eq:8xGrowingLos}{0.92}
\vspace{-6pt}
\resizeEq{
\mathcal{L}_{G_{1,8\times}}(D_{1t,8\times},D_{1s,8\times},G_{1,8\times})=&-\myavg_n[D_{1t,8\times}\Big(\widetilde{G_{1,8\times}}_{\mathcal{A}}\Big(\widetilde{X}\Big)\Big)]-\myavg_m[D_{1s,8\times}(x,{G_{1,8\times}}(x))] \\&+\lambda_{L1}\myavg_m[\vert\vert(G_{1,8\times}(x)-y)\vert\vert_1], \nonumber}{eq:8xGrowingLoss}{0.99}
with $\hat{y}=(1-R)\cdot G_{1,8\times}(x) + R\cdot y$, $\hat{Y}_{\mathcal{A}}=(1-R)\cdot \widetilde{G_{1,8\times}}_{\mathcal{A}}(\widetilde{X}) + R \cdot \widetilde{Y}_{\mathcal{A}}$, and $R\in[0,1]$, \MaxTodoNew{which is randomly sampled}. $n$ and $m$ denote the mini-batch size\marie{s} used for training the temporal and spatial discriminator and are set to $15$ and $16$, respectively. 
We use $\lambda_{W}=10$ and $\lambda_{L1}=20$ \marie{as scaling factors for the loss terms}.
The networks are trained by using Adam~\cite{Adam} with learning rate $\eta=0.0005$, $\beta_1 = 0.0$, $\beta_2 = 0.99$\marie{,} and $\epsilon=10^{-8}$. \MaxTodoNew{\marie{\myrefeq{eq:8xGrowingLos}} is used instead of \myrefeq{eq:4xlossfunction1} and \myrefeq{eq:4xlossfunction} when training $G_{1,8\times}$, $D_{1s,8\times}$\marie{,} and $D_{1t,8\times}$. Note that we do not employ the feature space loss since the training process stays stable \marie{by using the WGAN-GP loss}.}

Training processes of the spatial discriminators with different losses are shown in \myreffig{fig:loss_comparison}.
Here, lower values indicate that the discriminator is more successful at differentiating between real and fake images. In a) and b), \MaxTodoNew{which employ a tempoGAN with the loss described in \myrefeq{eq:4xlossfunction1}\marie{,} (\ref{eq:4xlossfunction}) and the least squares GAN loss \cite{LSGAN} (LSGAN)}, respectively, it becomes apparent that the classification task is too easy, especially after fading in new layers. Since the losses continue to \marie{drop} for tempoGAN and LSGAN which means that the spatial discriminators are growing too strong too quickly, the gradients for the discriminators will become more and more ineffective, which in turn leads to the generator receiving unreliable guidance \cite{trainingGANs}. 
In contrast, the stability of the WGAN with gradient penalty in c) is \marie{barely} influenced by the blending process, and can recover quickly from the curriculum updates.

\section{Data Generation}

The data used for training is generated by a stable fluids solver \cite{stam1999stable} with MacCormack advection and MiC-preconditioned CG solver in mantaflow \cite{mantaflow}. 
Overall, we use $20$ 3D simulations with $120$ frames each. For the setup, we initialize between $3$ and $12$ random density and velocity inflow areas combined with a randomized buoyancy force between $(0,0,0)$ and $(0,3\cdot10^{-4},0)$.
The LR volume is of size $64^3$, whereas the HR \MaxTodoNew{references are of size} $256^3$ for the tempoGAN $4\times$ up-scaling case, and \MaxTodoNew{of size }$512^3$ for the progressively growing GAN $8\times$ up-scaling case. The inputs $x$ are generated by down-scaling the reference volumes $y$. When loading and converting the data to slices, we remove slices with average density below the threshold 0.005.. 
For the progressively growing version, we also generate the intermediate resolution frames of size $128^3$ and $256^3$ via down-scaling the references.
We apply physical data augmentation methods in form of scaling and 90-degree rotations \cite{xie2018tempogan}. When modifying an input density field\marie{,} we also have to adjust the velocity field accordingly. 

\begin{figure}[h]
    \centering
\begin{overpic}[width=0.48\linewidth,tics=10,clip,trim={ 360px, 160px, 20px, 270px}]{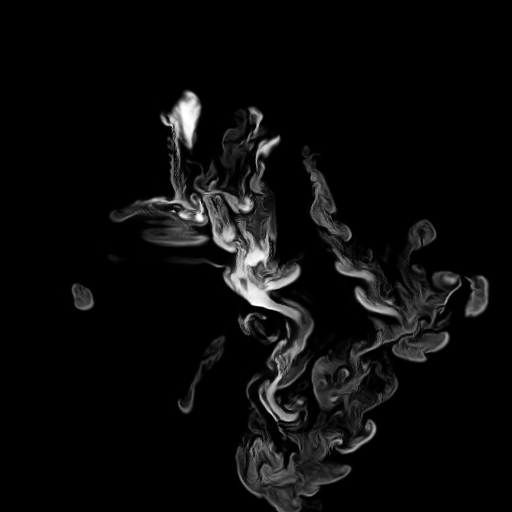}
 \put (3,4) {\textcolor{white}{$a)$}}
 \end{overpic} 
\begin{overpic}[width=0.48\linewidth,tics=10,clip,trim={ 360px, 160px, 20px, 270px}]{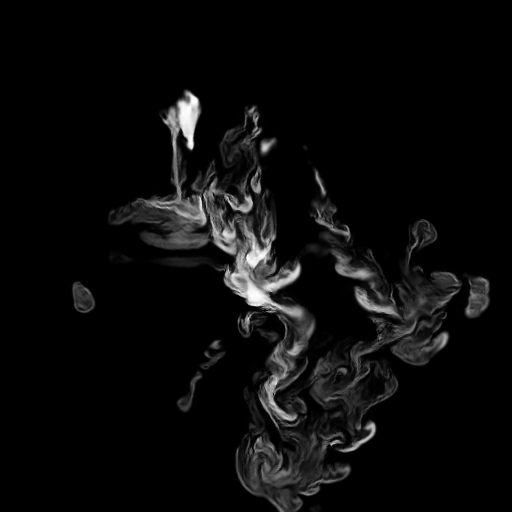}
 \put (3,4) {\textcolor{white}{$b)$}}
 \end{overpic} 
    \caption{\small {Comparison of a) using 90 degree rotations and flipping of the data as an additional data augmentation and b) \marie{only} using scaling.}}
    \label{fig:dataaugm}
\end{figure}

While fluids in general exhibit rotational invariance, we target phenomena with buoyancy which confines the invariance to rotations around the axis of gravity.
To train for shift invariance, we cut out tile pairs using randomized offsets before applying any augmentation. The size of the tiles after augmentation is $16^2$ for our inputs, and $(16\cdot j)^2$ for the targets of the current stage with $j$ being the up-scaling factor. 
Overall, data augmentation is crucial for successful training runs, as illustrated with an example in \myreffig{fig:dataaugm}. More varied training data \marie{encourages} the generation of more coherent densities with sharper details.

\begin{figure}
    \centering
\begin{overpic}[width=0.95\linewidth,tics=10,clip,trim={  25px, 30px, 20px, 260px}]{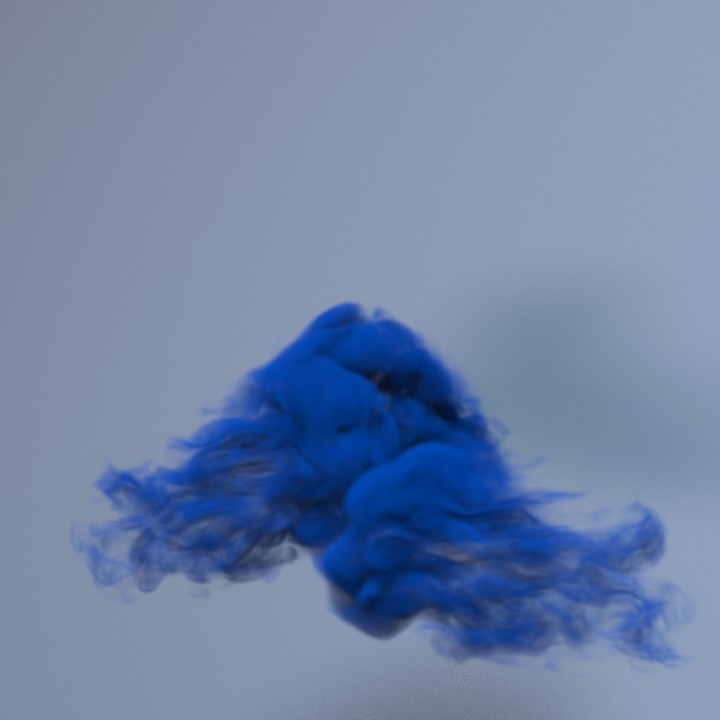}
 \put (3,4) {\textcolor{white}{$G_{2,8\times}$}}
 
 \put (2,45) {\begin{overpic}[width=0.27\linewidth,tics=10,clip,trim={  45px, 30px, 30px, 280px}]{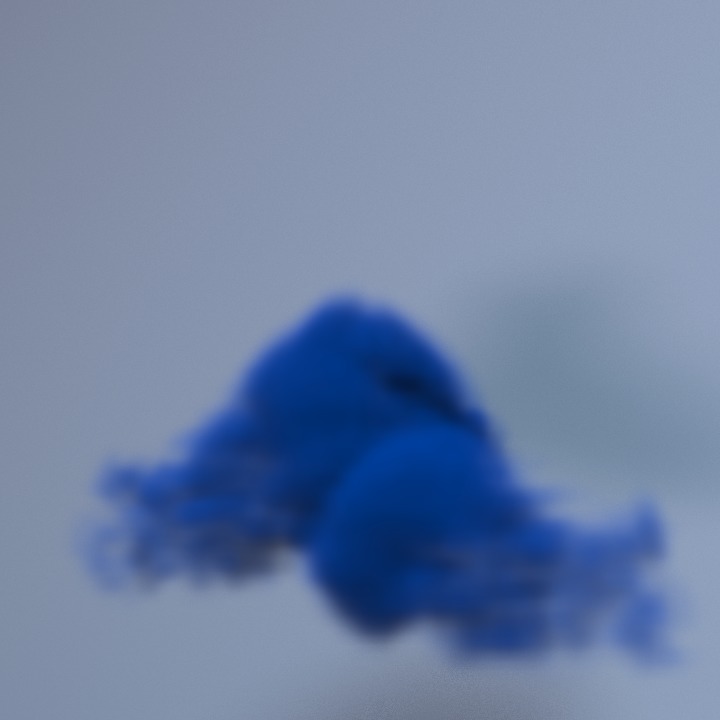}
 \put (3,4) {\small\textcolor{white}{$X$}}
 \end{overpic}}
 \end{overpic} 
    \vspace{-6pt}
    \caption{{\small Compared to the training data, this colliding smoke scene contains very different motions. Based on an input \marie{of size} $\mathbf{50^3}$, the network generates large amounts of realistic detail.}}
    \label{fig:out8x2}\vspace{-6pt}
\end{figure}

\section{Results and Discussion}

In the following, we demonstrate that our method can generate realistic outputs in a variety of flow settings, and that our network generalizes to a large class of buoyant smoke motions. For all the following scenes, please also consider the accompanying video which contains the full sequences in motion.

\begin{figure}[t]
    \centering
\begin{overpic}[width=0.95\linewidth,tics=10,clip,trim={  0px, 20px, 350px, 20px}]{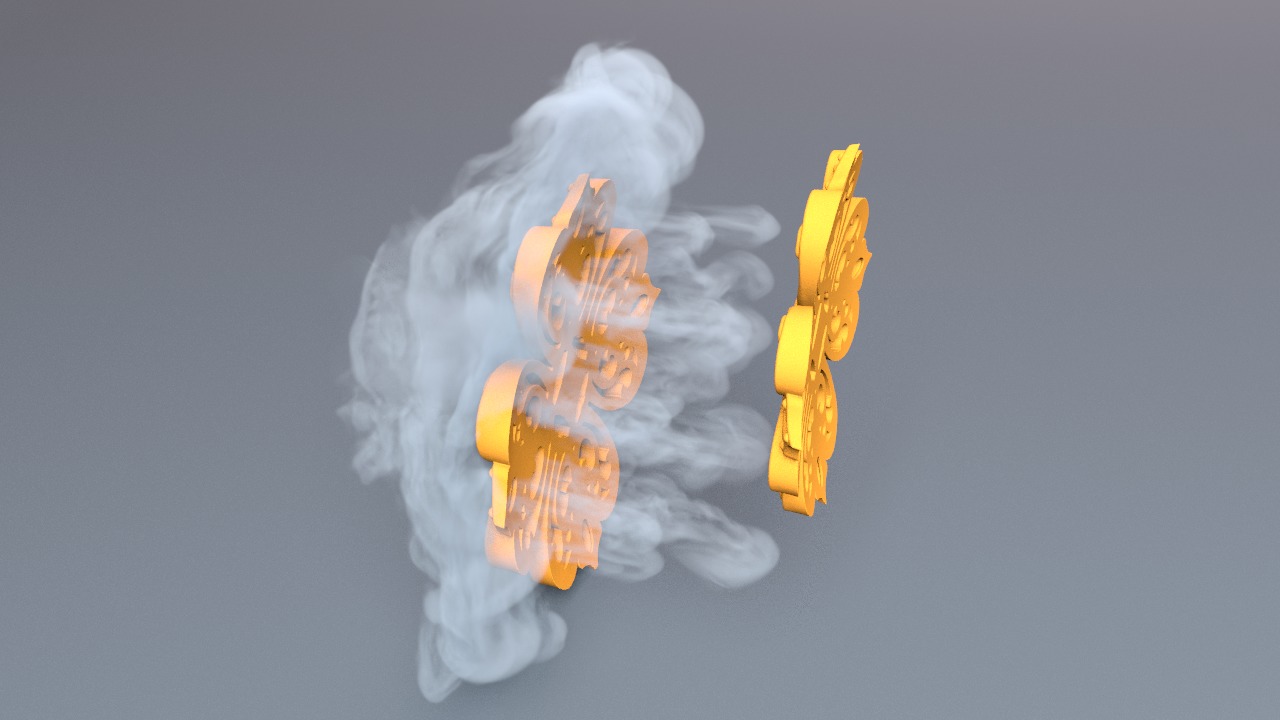}
 \put (3,4) {\textcolor{white}{$G_{2,8\times}$}}
 
 \put (3,30) {\begin{overpic}[width=0.28\linewidth,tics=10,clip,trim={  300px, 20px, 350px, 20px}]{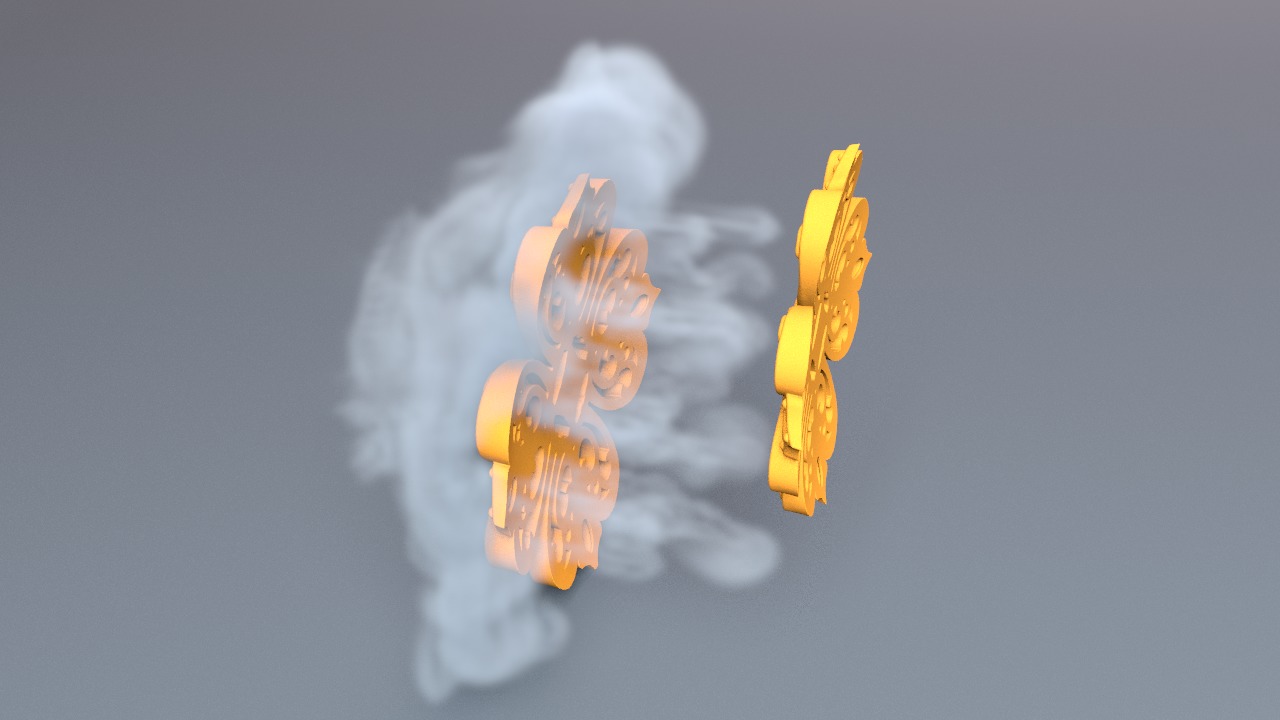}
 \put (3,4) {\textcolor{white}{$X$}}
 \end{overpic}}
 \end{overpic} 
    \caption{{\small Even when applied to fluid-obstacle simulations\marie{,} our algorithm reproduce\marie{s} the thin structures around objects. The resulting volume is of size $\mathbf{800^3}$. This simulation is the same as shown in \myreffig{fig:teaser}. Note that the network was only trained on buoyant smoke data.}}
    \label{fig:out8x1}
\end{figure}

In \myreffig{fig:out8x2}, a scene with three colliding smoke plumes is shown. The input of size $50^3$ is transformed into an output of $400^3$ by our network. The smooth streaks in the input are successfully sharpened and refined by our multi-pass GAN.
\nilsNew{In addition, the generated volumes are temporally coherent along all spatial dimensions.}
\myreffig{fig:out8x1} on the other hand shows a scene of fine smoke filaments interacting with a complex obstacle in the flow.
The underlying simulation is of size $100^3$, while the generated output has a resolution of $800^3$. Even though the training data did not contain any obstacles, our network manages to create realistic and detailed volumetric smoke effects.

\begin{figure}
    \centering
\begin{overpic}[width=0.44\linewidth,angle=180]{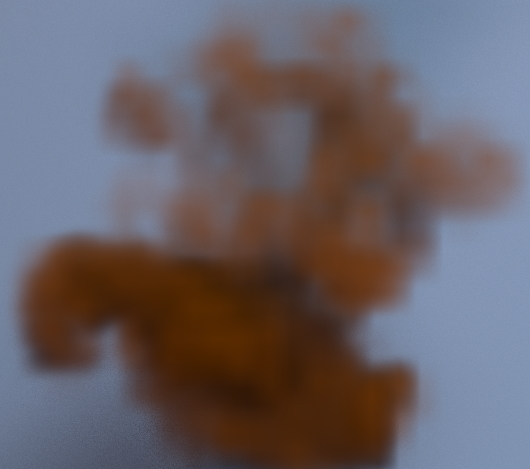}
 \put (3,4) {\textcolor{white}{$32^3$}}
 \end{overpic} 
\begin{overpic}[width=0.44\linewidth,angle=180]{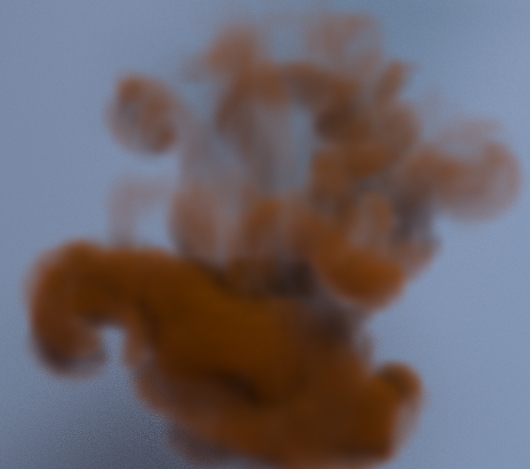}
 \put (3,4) {\textcolor{white}{$64^3$}}
 \end{overpic}  \\
\centering
\begin{overpic}[width=0.44\linewidth,angle=180]{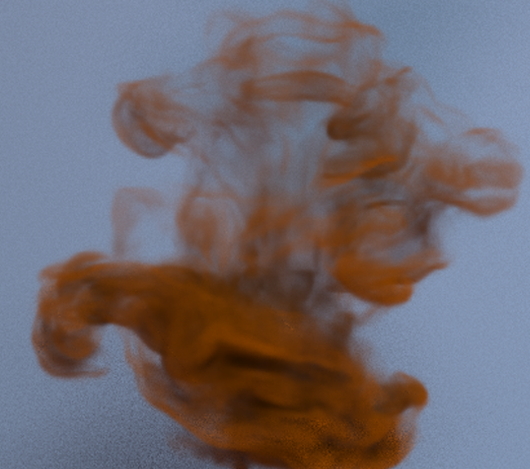}
 \put (3,5) {\textcolor{white}{$8\times_{32^3}$}}
 \end{overpic} 
\begin{overpic}[width=0.44\linewidth,angle=180]{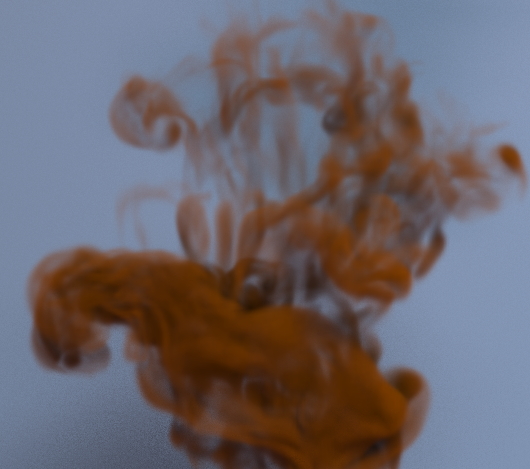}
 \put (3,4) {\textcolor{white}{$4\times$}}
 \end{overpic} \\
\begin{overpic}[width=0.44\linewidth,angle=180]{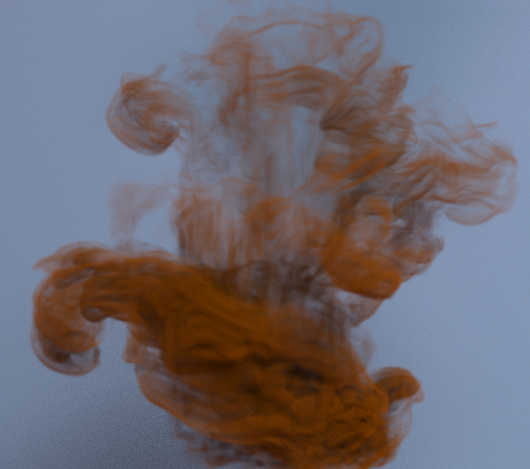}
 \put (3,5) {\textcolor{white}{$8\times_{64^3}$}}
 \end{overpic}
\begin{overpic}[width=0.44\linewidth,angle=180]{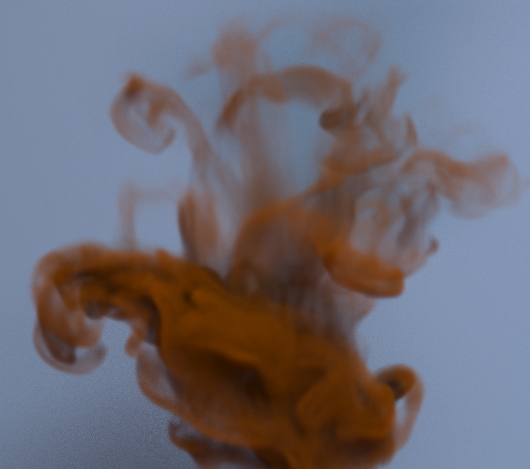}
 \put (3,4) {\textcolor{white}{\cite{xie2018tempogan}}}
 \end{overpic}
    \caption{{\small Comparison of $\mathbf{4\times}$ multi-pass GAN, \marie{the} $\mathbf{8\times}$ progressively growing version and the 3D tempoGAN, all at the same final resolution. Before applying the $\mathbf{8\times}$ model, the LR input was down-scaled once more for $\mathbf{8\times_{32^3}}$. Originally\marie{,} the simulation was of size $\mathbf{64^3}$. In addition, the lower left image show\marie{s} the result of the $\mathbf{8\times}$ model for a $\mathbf{64^3}$ input. The generated volume contains even sharper structures. }}
    \label{fig:4xvs8x}
\end{figure}

\begin{figure}[t]
    \centering
\begin{overpic}[width=0.235\linewidth,tics=10,clip,trim={  96px, 120px, 95px, 50px}]{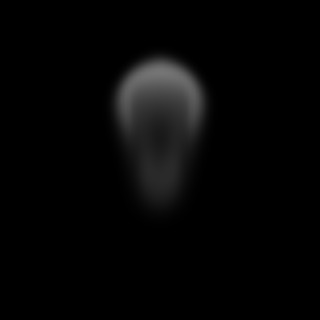}
 \put (3,4) {\small\textcolor{white}{$xy$}}
 \end{overpic}
\begin{overpic}[width=0.235\linewidth,tics=10,clip,trim={  30px, 60px, 100px, 40px}]{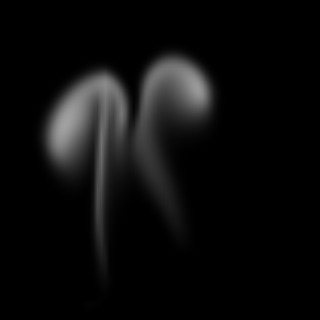}
 \put (3,4) {\small\textcolor{white}{$yz$}}
 \end{overpic}
\begin{overpic}[width=0.235\linewidth,tics=10,clip,trim={  96px, 120px, 95px, 50px}]{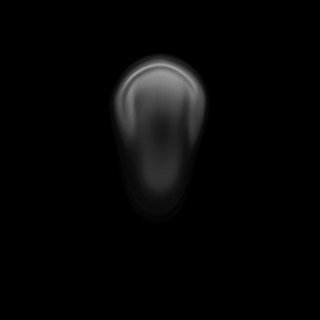}
 \put (3,4) {\small\textcolor{white}{$xy$}}
 \end{overpic}
\begin{overpic}[width=0.235\linewidth,tics=10,clip,trim={  30px, 60px, 100px, 40px}]{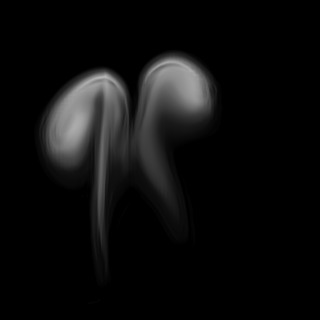}
 \put (3,4) {\small\textcolor{white}{$yz$}}
 \end{overpic}
\begin{overpic}[width=0.48\linewidth,tics=10,clip,trim={  450px, 60px, 300px, 60px}]{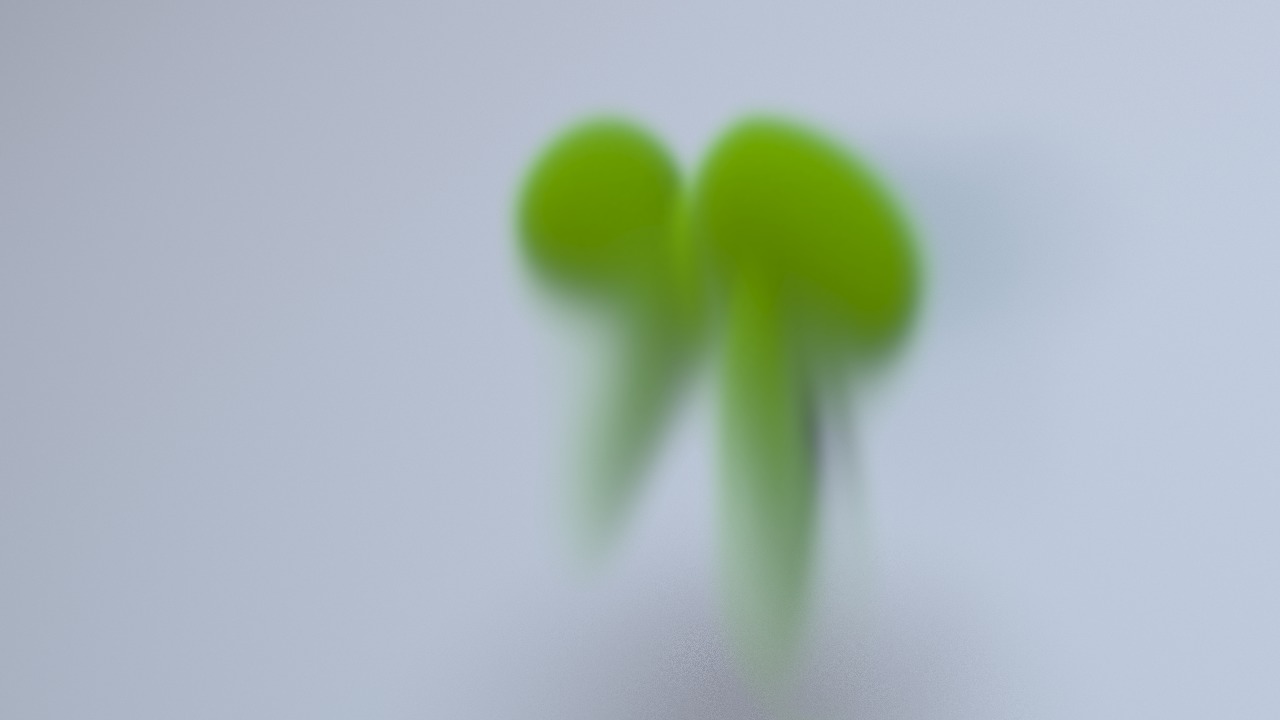}
 \put (3,4) {\small\textcolor{white}{$X$}}
 \end{overpic}
\begin{overpic}[width=0.48\linewidth,tics=10,clip,trim={  450px, 60px, 300px, 60px}]{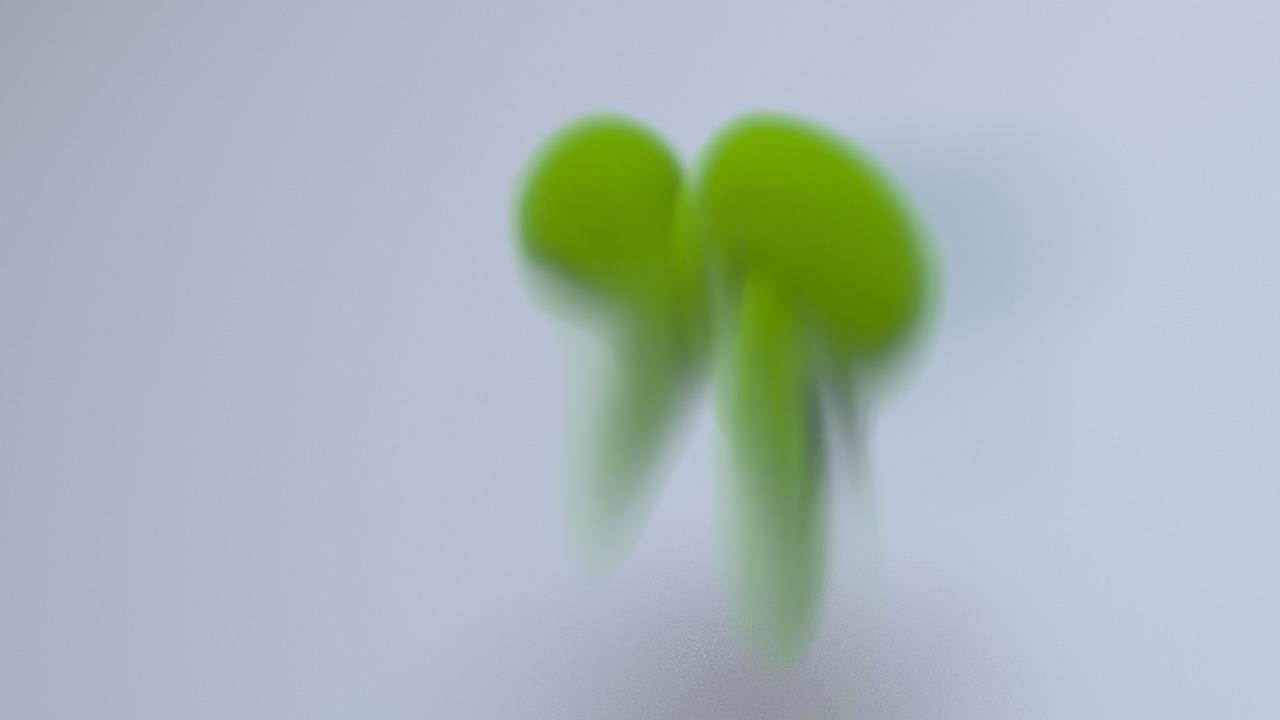}
 \put (3,4) {\textcolor{white}{$G_2$}}
 \end{overpic} 
    \caption{\MaxTodoNew{{\small Applying the algorithm to a smooth input simulation with velocities of low magnitude results in the generation of barely any additional details. This is the desired outcome: the networks  preserves the rather smooth regions of the input.}}}
    \label{fig:out4xsmooth}
\end{figure}

\begin{figure}
    \centering
    \vspace{-6pt}
\begin{overpic}[width=0.3\linewidth,tics=10,clip,trim={ 790px, 295px, 230px, 175px}]{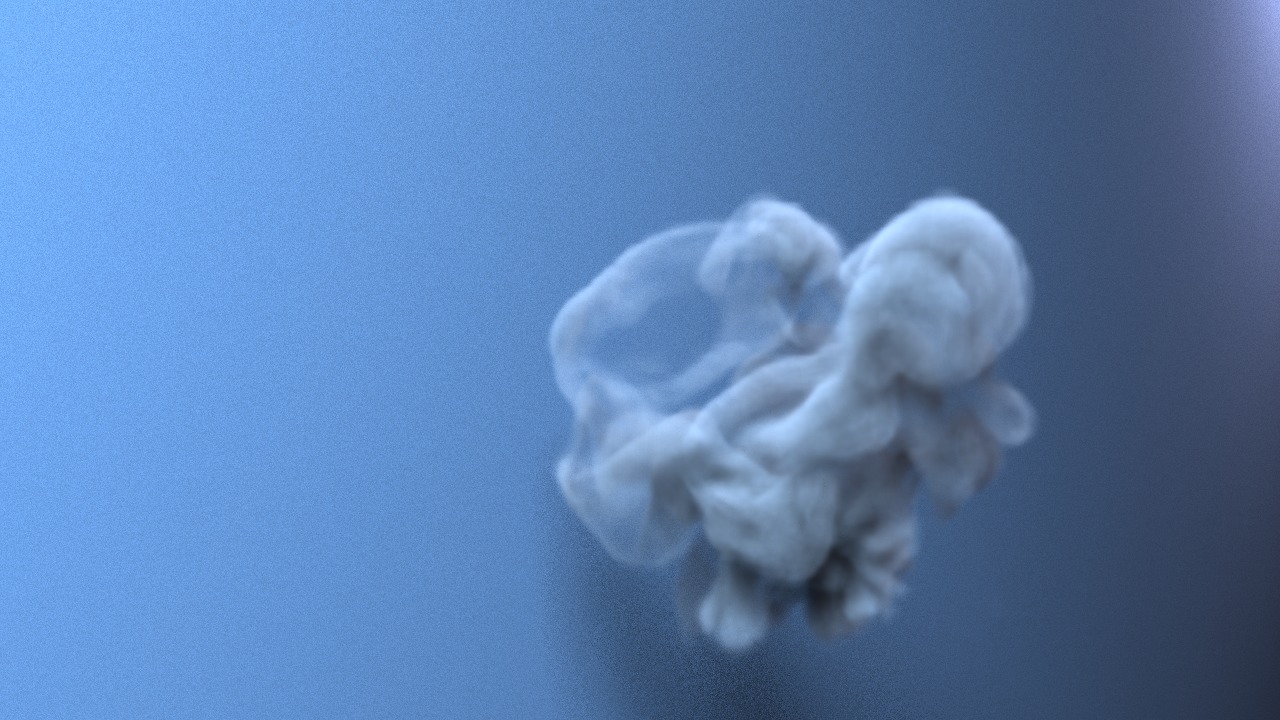}
 \put (3,4) {\textcolor{white}{$AVG$}}
 \end{overpic} 
\begin{overpic}[width=0.3\linewidth,tics=10,clip,trim={ 790px, 295px, 230px, 175px}]{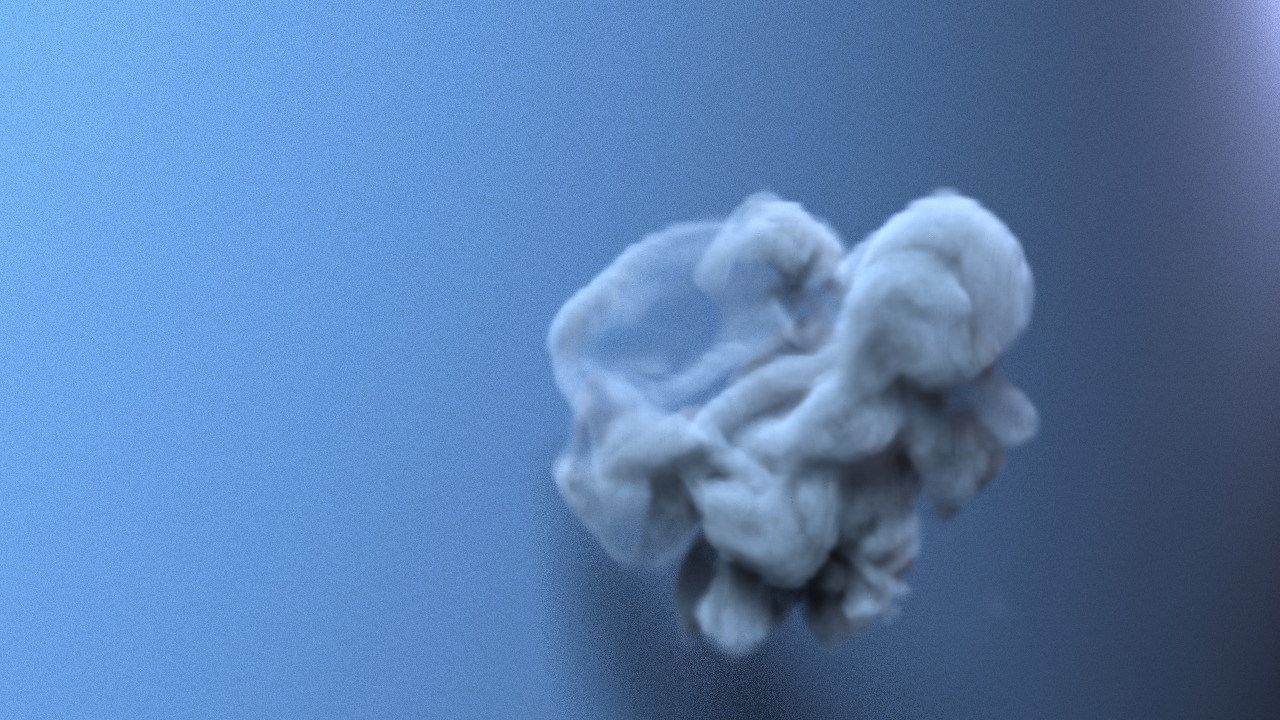}
 \put (3,4) {\textcolor{white}{$MAX$}}
 \end{overpic} 
\begin{overpic}[width=0.3\linewidth,tics=10,clip,trim={ 790px, 295px, 230px, 175px}]{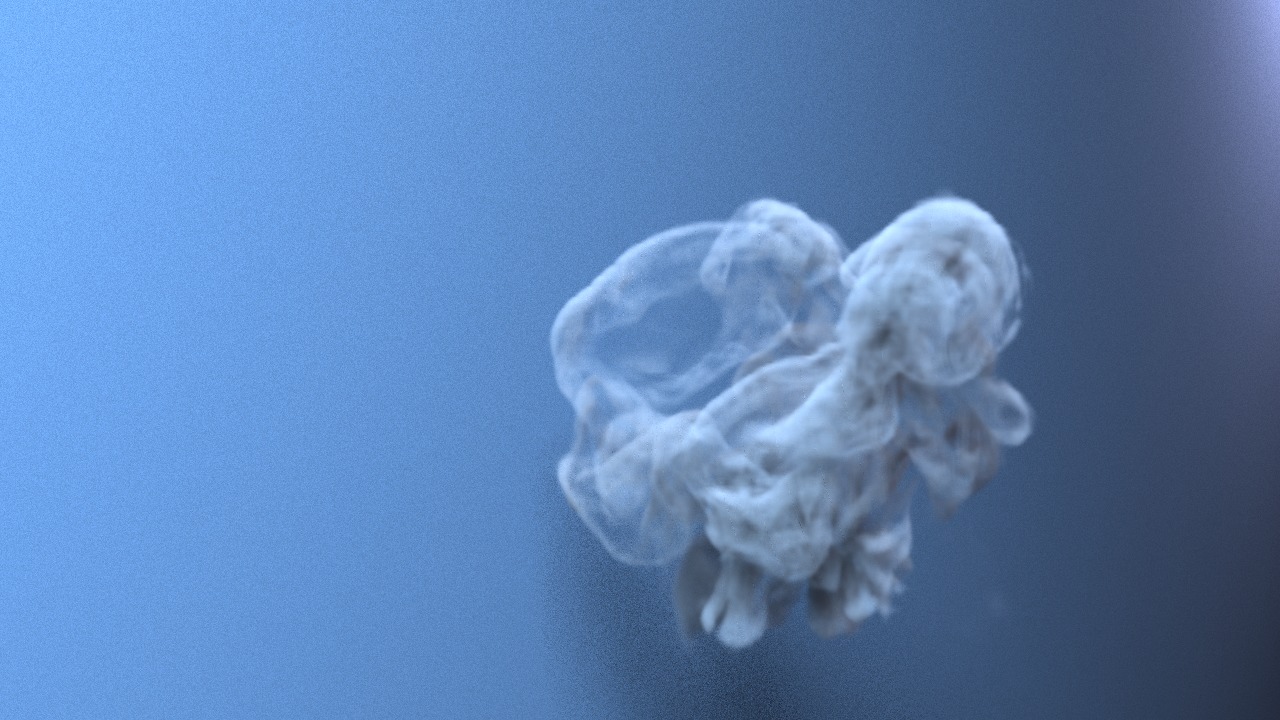}
 \put (3,4) {\textcolor{white}{$RES$}}
 \end{overpic} \\
\centering
\begin{overpic}[width=0.3\linewidth,tics=10,clip,trim={ 790px, 295px, 230px, 175px}]{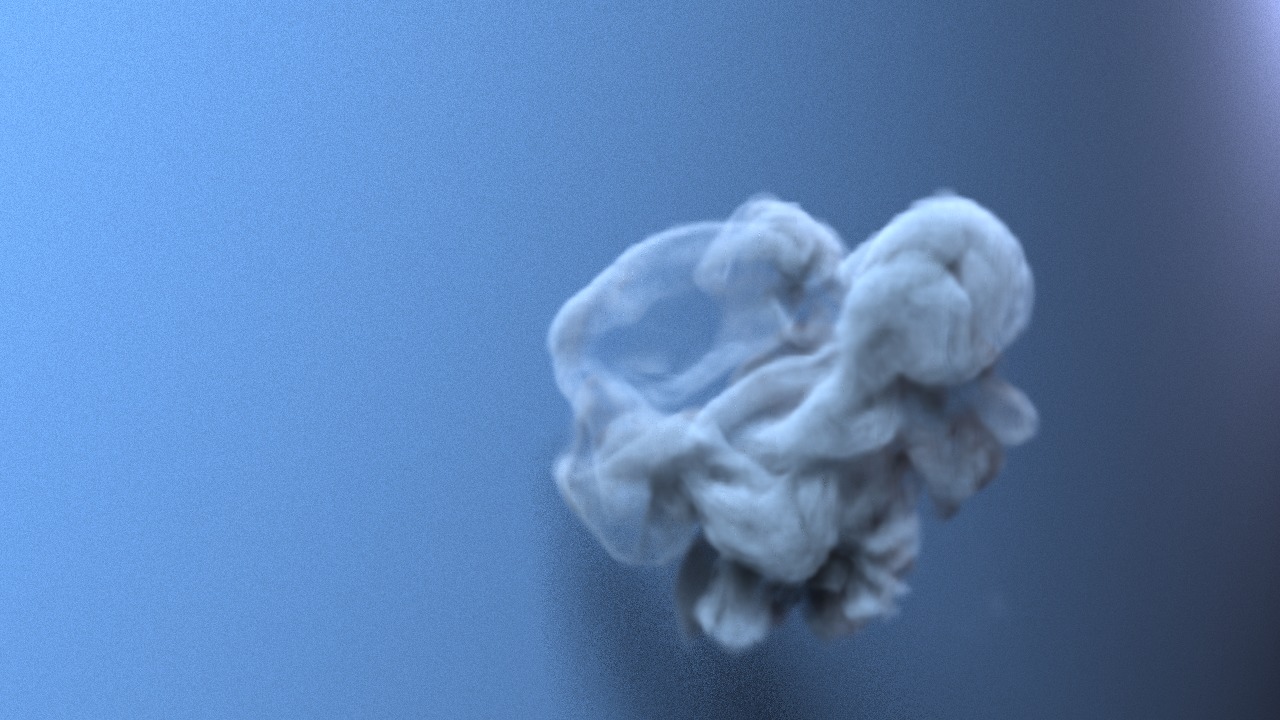}
 \put (3,4) {\textcolor{white}{$CRES$}}
 \end{overpic} 
\begin{overpic}[width=0.3\linewidth,tics=10,clip,trim={ 790px, 295px, 230px, 175px}]{2ndnn_01}
 \put (3,4) {\textcolor{white}{$G_2$}}
 \end{overpic} 
\begin{overpic}[width=0.3\linewidth,tics=10,clip,trim={ 790px, 295px, 230px, 175px}]{high_01}
 \put (3,4) {\textcolor{white}{$Y$}}
 \end{overpic} 
    \caption{\small {Different combination techniques of multiple volumes which were up-sampled along different directions. In comparison to this, our method produces sharper edges and no stripe artifacts in any dimension.}}
    \label{fig:naive1}
\vspace{-6pt}
\end{figure}

\subsection{Evaluation}

In addition, it is interesting to compare the performance of our models trained for different up-sampling factors with outputs at the same target resolution. To compare the $8\times$ model with the $4\times$ one, we apply the former one to a simulation of size $32^3$, which was generated by down-sampling the actual simulation of size $64^3$. 
This way, both models generate a final volume of size $256^3$.
\myreffig{fig:4xvs8x} shows renderings of the outputs. Here it becomes apparent that our $8\times$ network manages to create even sharper and more pronounced details than the $4\times$ version, despite starting from a coarser input. \MaxTodoNew{For this simulation\marie{,} we \marie{used the time step $\Delta t_o=0.25$ instead of $\Delta t_t=0.5$ as for training our networks, with $\Delta t_t$ and $\Delta t_o$ denoting the training and output time step, respectively}. Most important\marie{ly,} the range of velocities of the input simulation should not exceed the range the network was trained on, otherwise artifacts \marie{occasionally} occur. \marie{One possibility is to} re-scale the velocities to match the training time step by multiplying with $\frac{\Delta t_t}{\Delta t_o}$ which we applied in the former example.}\par
\MaxTodoNew{As \marie{shown} in \myreffig{fig:out4xsmooth}, applying the $4\times$ networks to a smooth LR simulation of size $80^3$ sharpens only very few areas, e.g., the area of the plume which faces the direction of the buoyancy force. \marie{As desired}, other parts such as the density inflow area 
are largely unchanged.
Notably, however, are the few new density gradients at certain boundaries of the smoke which might be undesirable when targeting a very smooth scenarios. Since our training data did not contain such smooth simulations, the results shown here could be improved by fine-tuning the networks further.}

As a simpler variant of our approach, we also evaluated 
applying the same network to all slices of a volume along all three dimensions, each of which was linearly up-sampled. This yields three volumes at full resolution which can be combined to obtain a single final output volume.
For the combination\marie{,} we have tested an averaging operation (AVG), taking the maximum per cell (MAX), or taking one of the volumes as a basis and then adding details by taking the difference between the the linearly up-sampled input and the output along another axis (RES). Th\marie{e} latter variant represents a residual transfer. Here, we also tested a variant that only used additive details to be transferred, i.e., the residual\marie{s} were clamped at zero (CRES).

As \marie{shown} in \myreffig{fig:naive1}, all of these simpler methods fail at producing sharp, round edges and typically generate staircasing artifacts. The (RES) version contains especially strong artifacts \marie{while} the (CRES) version does not perform much better. In comparison, our approach yields smooth and detailed outputs. Thus, training specialized networks for multiple passes is preferable over \marie{re-using} a single network. The additional work to train $G_2$ for a second refinement pass pays off in terms of quality of the generated structures.

\subsection{Performance}\label{sec:performance}

Despite making it possible to achieve large up-scaling factors, our method
also reduces training time.
Training our $4\times$ multi-pass network took approximately $3$ days for the first generator and $2$ days for the second. This is almost twice as fast as training a 3D model reported by previous work \cite{xie2018tempogan}. 
We additionally only employed a single GTX 1080 Ti instead of two GPUs. 
Training the progressively growing network for an $8\times$ up-scaling took about $8$ and $5$ days for the first and second generator network, respectively. As this scale is not feasible \marie{with} previous work, we cannot compare training times for the $8\times$ case.

Similar to previous work, the memory available in current GPUs can be a bottleneck when applying the trained networks to new input, and can make it necessary to subdivide the inputs into tiles.
\marie{Regarding} our implementation\marie{,} we \marie{are able to} deal with full slices of up to approximately $256^3$. Therefore, we usually do not need to apply tiling. 
For larger volumes, the tiling process requires an additional overlap
for the tiles ($4$ \MaxTodoNew{LR} cells for $G_1$ and $16$ \MaxTodoNew{HR} cells for $G_2$).
It takes around $10.14$ seconds to apply our $4\times$ network to a single frame of size $64^3$ to up-scale the volume to $256^3$. Applying $G_{1,8\times}$ to a volume of size $64^3$ takes $7.58$ seconds while refining it with $G_{2,8\times}$ takes an additional $52.07$ seconds. 
\YouAdd{We compared our multi-pass GAN with a regular CPU-based solver using CG or multi-grid pressure solvers which are shown in ~\myreftab{tab:performance}. According to the CFL condition, we applied \marie{a} $\frac{1}{f}$ smaller time step (where the \marie{up-scaling factor} $f$
is $4$ or $8$ in our case) for stability and \marie{con}vergence. E.g., if we generate data \MaxTodoNew{of the same time length} as \MaxTodoNew{with the} $8\times$ multi-pass GAN, we run $8\times$ more iterations with a regular or multi-grid fluid solver with a $\frac{1}{8}$ time step}. Note however, that this solver is CPU-based and therefore it is difficult to compare the actual timings. Applying our method scales linearly with the number of cells of the simulation, whereas fluid solvers with CG-based pressure projections typically scale super-linearly. \YouAdd{Besides, for a regular solver, simulation time noticeably \marie{rises} when the smoke volume increases in later frames while the content does not affect generation time for the multi-pass GAN. \marie{In} ~\myreftab{tab:performance}, we clearly see that the multi-pass GAN is significantly faster than regular and multi-grid fluid solvers when generating data \nils{for a given number of frames}.}
\setlength{\tabcolsep}{2pt}
\begin{table}[]
\YouAdd{
\caption{\YouAdd{{\footnotesize{Evaluation of Performance (avg. for 100 frames)}}}}
\label{tab:performance}
\begin{tabular}{|c|c|c|c|c|c|c|}
\hline
                                                                                      & \multicolumn{2}{c|}{\begin{tabular}[c]{@{}c@{}}Regular\\ solver\end{tabular}} & \multicolumn{2}{c|}{\begin{tabular}[c]{@{}c@{}}Multi-grid\\ solver\end{tabular}} & \multicolumn{2}{c|}{\begin{tabular}[c]{@{}c@{}}Multi-pass\\ GAN\end{tabular}} \\ \hline
Resolution                                                                            & 256                                   & 512                                   & 256                                     & 512                                    & 256                                   & 512                                   \\ \hline
\begin{tabular}[c]{@{}c@{}}Computation time (s)\end{tabular} & 116.90                                      & 1376.54                                       & 41.90                                       &  463.81                                      & 10.14                                      & 59.65                                       \\ \hline
\end{tabular}}
\end{table}

\subsection{Limitations and Outlook}\label{sec:limits}

A first limitations of our approach is that it requires multiple passes
over the full HR volume (two in our case). However, in practice\marie{,} these
passes often result in speed-ups compared to networks that have to process the
full data set at once. E.g., the tempoGAN architecture relies on up to $128$ 
latent features per volumetric degree of freedom. These intermediate representations
have to be stored on the GPU and quickly fill up the
memory capacities of current GPU architectures. Hence, larger output volumes will 
require tiling of the inputs and \marie{induces increased workloads due to these} ghost layers.
In contrast, our multi-pass networks only build feature spaces for slices of reduced 
dimensionality that typically can be processed without tiling.

\begin{figure}[t]
    \centering
    \vspace{-3pt}
\begin{overpic}[width=0.49\linewidth, height=0.4\linewidth,tics=10,clip,trim={  445px, 395px, 135px, 225px}]{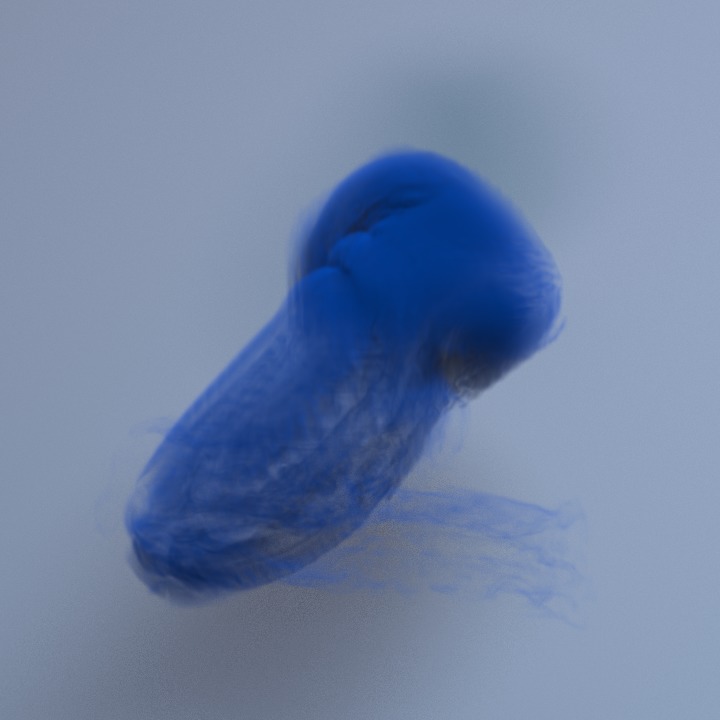}
 \put (3,4) {\textcolor{white}{$a)$}}
 \end{overpic} 
\begin{overpic}[width=0.49\linewidth, height=0.4\linewidth,tics=10,clip,trim={  60px, 240px, 360px, 240px}]{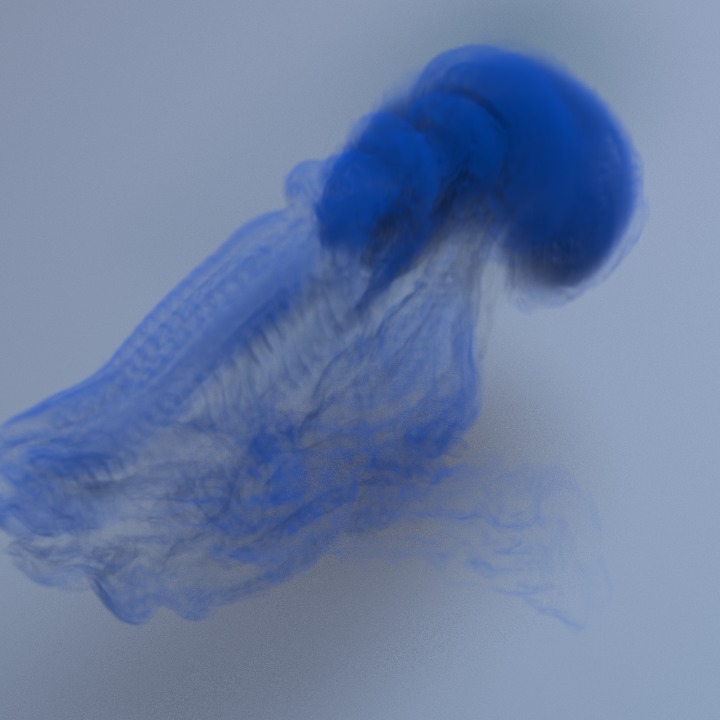}
 \put (3,4) {\textcolor{white}{$b)$}}
 \end{overpic} 
 	\vspace{-9pt}
    \caption{\nilsNew{Examples of patterns that our $8\times$ network generate\marie{s} in unfavorable situations.}
    }
    \label{fig:AAartifacts}
\end{figure}

Our $8\times$ network in its current form can lead to 
artifacts caused by the initial LR sampling. This typically only happens 
in rare situations, two examples of which are shown in \myreffig{fig:AAartifacts}.
Here, the network seems to reinforce steps from the input data.
These most likely stem from the difficult inference task to generate a $512\times$ larger output. 
As we have not observed these artifacts in our $4\times$ versions, they potentially could 
be alleviated by larger and more varied training data sets.

\section{Conclusion}

We have presented a first multi-pass GAN method to achieve volumetric generative networks that can up-sample 3D spatio-temporal smoke effects by factors of eight. We have demonstrated this for the tempoGAN setting and demonstrated that our approach can be extended to curriculum learning and \marie{g}rowing GANs. Most importantly, our method reduces the number of weights that need to be trained at once, which yields shorter and more robust training runs. While we have focused on single-phase flow effects in 3D, our method is potentially applicable to all kinds of high-dimensional field data, e.g., it will be highly interesting to apply our method to 4D space-time data sets.

\section{Acknowledgments}
This work was funded by the ERC Starting Grant realFlow (StG-2015-
637014). We would like to thank Marie-Lena Eckert for paper proofreading.

\appendix
\section*{APPENDIX}
\setcounter{section}{0}

\section{Network architecture}
\label{appendix:A}

\small

\setlength{\tabcolsep}{2pt}
\begin{table}[H]
	\vspace{-6pt}
	\parbox{.41\linewidth}{
		\caption{{\footnotesize{Architecture of $G_{1,8\times}$}}}
		\vspace{-6pt}
		\label{tab:g1archetecture}
		\begin{tabular}{cc}
			\hline
			Layer            & Output Shape \\ \hline \\
			LR Input         & 16$\times$16$\times$4      \\
			ResBlock. 3$\times$3    & 16$\times$16$\times$16     \\
			ResBlock. 3$\times$3    & 16$\times$16$\times$64     \\ \hline
			Avg-Depool       & 32$\times$32$\times$64     \\
			ResBlock. 3$\times$3    & 32$\times$32$\times$128    \\
			ResBlock. 3$\times$3    & 32$\times$32$\times$64     \\\hline
			Avg-Depool       & 64$\times$64$\times$64     \\        
			ResBlock. 3$\times$3    & 64$\times$64$\times$64     \\
			ResBlock. 3$\times$3    & 64$\times$64$\times$32     \\ \hline
			Avg-Depool       & 128$\times$128$\times$32   \\
			ResBlock. 3$\times$3    & 128$\times$128$\times$32   \\
			ResBlock. 3$\times$3    & 128$\times$128$\times$16   \\
			Conv. 1$\times$1        & 128$\times$128$\times$1    \\ \hline
			total parameters & 550k         \\ \hline
		\end{tabular}
	}
	\hfill
	\parbox{.52\linewidth}{
		\caption{{\scriptsize {Architecture of $D_{1s,8\times}$ \& $D_{1t,8\times}$}}}
		\vspace{-6pt}
		\label{tab:d1archetecture}
		\begin{tabular}{cc}
			\hline
			Layer            & Output Shape     \\ \hline
			HR Input         & 128$\times$128$\times$\{2, 3\} \\
			Conv. 1$\times$1        & 128$\times$128$\times$32       \\
			Conv. 3$\times$3        & 128$\times$128$\times$32       \\
			Conv. 3$\times$3        & 128$\times$128$\times$64       \\\hline
			Avg-Pool         & 64$\times$64$\times$64        \\ 
			Conv. 3$\times$3        & 64$\times$64$\times$64         \\
			Conv. 3$\times$3        & 64$\times$64$\times$128        \\\hline
			Avg-Pool         & 32$\times$32$\times$128        \\ 
			Conv. 3$\times$3        & 32$\times$32$\times$128        \\
			Conv. 3$\times$3        & 32$\times$32$\times$128        \\\hline
			Avg-Pool         & 16$\times$16$\times$128        \\ 
			Conv. 3$\times$3        & 16$\times$16$\times$32         \\
			Conv. 3$\times$3        & 16$\times$16$\times$4          \\
			Flatten \& FC    & 1$\times$1$\times$1            \\ \hline
			total parameters & 470k             \\ \hline
		\end{tabular}
	}
\vspace{-6pt}
\end{table}
\begin{table}[H]
	\parbox{.41\linewidth}{
		\caption{{\footnotesize{Architecture of $G_{2,8\times}$}}}
		\label{tab:g2archetecture}
		\begin{tabular}{cc}
			\hline
			Layer            & Output Shape   \\ \hline \\
			LR Input         & 16$\times$16$\times k_{i}$ \\
			ResBlock. 5$\times$5    & 64$\times$64$\times$12       \\
			ResBlock. 5$\times$5    & 64$\times$64$\times$48       \\ \hline
			ResBlock. 5$\times$5    & 64$\times$64$\times$96       \\
			ResBlock. 5$\times$5    & 64$\times$64$\times$48       \\ \hline
			ResBlock. 5$\times$5    & 64$\times$64$\times$48       \\
			ResBlock. 5$\times$5    & 64$\times$64$\times$24       \\ \hline
			ResBlock. 5$\times$5    & 64$\times$64$\times$24       \\
			ResBlock. 5$\times$5    & 64$\times$64$\times$12       \\
			Conv. 1x1        & 64$\times$64$\times$1      \\ \hline
			total parameters & 774k           \\ \hline
		\end{tabular}
	}
	\hfill
	\parbox{.52\linewidth}{
		\caption{\scriptsize {Architecture of $D_{2s,8\times}$ \& $D_{2t,8\times}$}}
		\label{tab:d2archetecture}
		\begin{tabular}{cc}
			\hline
			Layer            & Output Shape     \\ \hline
			HR Input         & 64$\times$64$\times$\{2, 3\} \\
			Conv. 1$\times$1        & 64$\times$64$\times$24       \\
			Conv. 5$\times$5        & 64$\times$64$\times$24       \\
			Conv. 5$\times$5        & 64$\times$64$\times$48       \\ \hline
			Conv. 5$\times$5        & 64$\times$64$\times$48         \\
			Conv. 5$\times$5        & 64$\times$64$\times$96        \\ \hline
			Conv. 5$\times$5        & 64$\times$64$\times$96        \\
			Conv. 5$\times$5        & 64$\times$64$\times$96        \\\hline
			Conv. 5$\times$5        & 64$\times$64$\times$32         \\
			Conv. 5$\times$5        & 64$\times$64$\times$4          \\
			Flatten \& FC    & 1$\times$1$\times$1            \\ \hline
			total parameters & 773k             \\ \hline
		\end{tabular}
	}
\end{table}

Architectures of $G_{1,8\times}$, $G_{2,8\times}$, $D_{s,8\times}$\marie{,} and $D_{t,8\times}$ are listed in ~\myreftab{tab:g1archetecture},~\myreftab{tab:d1archetecture},~\myreftab{tab:g2archetecture}\marie{,} and~\myreftab{tab:d2archetecture}, respectively. ReLU is used for the generator and an additional pixel-wise normalization layer is added after every convolutional layer. Each convolutional layer in the discriminator uses leaky ReLU as an activation function. The input \marie{to} the spatial discriminator consists of the up-scaled LR and HR density fields, whereas the input \marie{to} the temporal one is composed of three advected frames. Input \marie{to} $G_{1,8\times}$ \marie{are} the LR density and velocity, whereas \marie{the input to} $G_{2,8\times}$ consists of LR density, velocity\marie{,} and the output of $G_{1,8\times}$. Note that the tile sizes for $G_{2,8\times}$ and $G_{1,8\times}$ are $64^2$ HR and $16^2$ LR pixels, respectively.

\section{Training parameters}
\label{appendix:B}
For an up-scaling factor of $8$, \marie{$3$ stages exist} in which new layers are faded in. 'blend iter.' describes the number of training iterations for $G_{1,8\times}$. The blending and stabilizing processes are applied in an alternating fashion: $120$k iterations for fading in, $120$k iterations for stabilizing, etc. This leads to $120k\cdot6=720k$ iterations overall. After finishing the progressive growing of the networks, we slowly decay the learning rate for 'decay iter.'.

\begin{table}[H]
\caption{Parameters for progressive growing}
\label{tab:gparams}
\begin{center}
\small{
\begin{tabular}{|c|c|c|c|c|c|}
\hline
Networks & blend  & decay  & training & time & batch \& \\ 
 & iter.  & iter.  & parameters &  & tile size \\ \hline
$G_{1,8\times}$, $D_{1s,8\times}$ & 120k & 160k & Adam: $\big[\eta=0.0005$, $\beta_1=0.0$, &  8&16,\\%\hline
\& $D_{1t,8\times}$ & & & $\beta_2=0.99$, $\epsilon=10^{-8}\big],$  &days & $16^2$\\
 & & & $\lambda_1=20, \lambda_W=10$ & & \\\hline
$G_{2,8\times}$, $D_{2s,8\times}$ & - & 600k & Adam: $\big[\eta=0.0005$, $\beta_1=0.0$,   &  5&16, \\%\hline
\& $D_{2t,8\times}$ & & &$\beta_2=0.99$, $\epsilon=10^{-8}\big]$, & days&$64^2$\\
 & & & $\lambda_1=20, \lambda_W=10$ & &\\\hline
\end{tabular}
}
\end{center}
\end{table}

\bibliographystyle{ACM-Reference-Format}
\bibliography{reference}


\begin{thebibliography}{48}

%%% the Web address.

\ifx \showCODEN    \undefined \def \showCODEN     #1{\unskip}     \fi
\ifx \showDOI      \undefined \def \showDOI       #1{#1}\fi
\ifx \showISBNx    \undefined \def \showISBNx     #1{\unskip}     \fi
\ifx \showISBNxiii \undefined \def \showISBNxiii  #1{\unskip}     \fi
\ifx \showISSN     \undefined \def \showISSN      #1{\unskip}     \fi
\ifx \showLCCN     \undefined \def \showLCCN      #1{\unskip}     \fi
\ifx \shownote     \undefined \def \shownote      #1{#1}          \fi
\ifx \showarticletitle \undefined \def \showarticletitle #1{#1}   \fi
\ifx \showURL      \undefined \def \showURL       {\relax}        \fi
% invisible to TeX
\providecommand\bibfield[2]{#2}
\providecommand\bibinfo[2]{#2}
\providecommand\natexlab[1]{#1}
\providecommand\showeprint[2][]{arXiv:#2}

\bibitem[\protect\citeauthoryear{Arjovsky and Bottou}{Arjovsky and
  Bottou}{2017}]%
        {trainingGANs}
\bibfield{author}{\bibinfo{person}{Mart{\'{\i}}n Arjovsky} {and}
  \bibinfo{person}{L{\'{e}}on Bottou}.} \bibinfo{year}{2017}\natexlab{}.
\newblock \showarticletitle{Towards Principled Methods for Training Generative
  Adversarial Networks}.
\newblock \bibinfo{journal}{\emph{CoRR}}  \bibinfo{volume}{abs/1701.04862}
  (\bibinfo{year}{2017}).
\newblock
\showeprint[arxiv]{1701.04862}
\urldef\tempurl%
\url{http://arxiv.org/abs/1701.04862}
\showURL{%
\tempurl}


\bibitem[\protect\citeauthoryear{Bako, Vogels, McWilliams, Meyer, Nov{\'a}k,
  Harvill, Sen, Derose, and Rousselle}{Bako et~al\mbox{.}}{2017}]%
        {bako2017kernel}
\bibfield{author}{\bibinfo{person}{Steve Bako}, \bibinfo{person}{Thijs Vogels},
  \bibinfo{person}{Brian McWilliams}, \bibinfo{person}{Mark Meyer},
  \bibinfo{person}{Jan Nov{\'a}k}, \bibinfo{person}{Alex Harvill},
  \bibinfo{person}{Pradeep Sen}, \bibinfo{person}{Tony Derose}, {and}
  \bibinfo{person}{Fabrice Rousselle}.} \bibinfo{year}{2017}\natexlab{}.
\newblock \showarticletitle{Kernel-predicting convolutional networks for
  denoising Monte Carlo renderings.}
\newblock \bibinfo{journal}{\emph{ACM Trans. Graph.}} \bibinfo{volume}{36},
  \bibinfo{number}{4} (\bibinfo{year}{2017}), \bibinfo{pages}{97--1}.
\newblock


\bibitem[\protect\citeauthoryear{Batty, Bertails, and Bridson}{Batty
  et~al\mbox{.}}{2007}]%
        {batty2007fast}
\bibfield{author}{\bibinfo{person}{Christopher Batty},
  \bibinfo{person}{Florence Bertails}, {and} \bibinfo{person}{Robert Bridson}.}
  \bibinfo{year}{2007}\natexlab{}.
\newblock \showarticletitle{A fast variational framework for accurate
  solid-fluid coupling}. In \bibinfo{booktitle}{\emph{ACM Transactions on
  Graphics (TOG)}}. ACM, \bibinfo{pages}{100}.
\newblock


\bibitem[\protect\citeauthoryear{Bhattacharjee and Das}{Bhattacharjee and
  Das}{2017}]%
        {bhattacharjee2017temporal}
\bibfield{author}{\bibinfo{person}{Prateep Bhattacharjee} {and}
  \bibinfo{person}{Sukhendu Das}.} \bibinfo{year}{2017}\natexlab{}.
\newblock \showarticletitle{Temporal coherency based criteria for predicting
  video frames using deep multi-stage generative adversarial networks}. In
  \bibinfo{booktitle}{\emph{Advances in Neural Information Processing
  Systems}}. \bibinfo{pages}{4268--4277}.
\newblock


\bibitem[\protect\citeauthoryear{Chaitanya, Kaplanyan, Schied, Salvi, Lefohn,
  Nowrouzezahrai, and Aila}{Chaitanya et~al\mbox{.}}{2017}]%
        {chaitanya2017interactive}
\bibfield{author}{\bibinfo{person}{Chakravarty R~Alla Chaitanya},
  \bibinfo{person}{Anton~S Kaplanyan}, \bibinfo{person}{Christoph Schied},
  \bibinfo{person}{Marco Salvi}, \bibinfo{person}{Aaron Lefohn},
  \bibinfo{person}{Derek Nowrouzezahrai}, {and} \bibinfo{person}{Timo Aila}.}
  \bibinfo{year}{2017}\natexlab{}.
\newblock \showarticletitle{Interactive reconstruction of Monte Carlo image
  sequences using a recurrent denoising autoencoder}.
\newblock \bibinfo{journal}{\emph{ACM Transactions on Graphics (TOG)}}
  \bibinfo{volume}{36}, \bibinfo{number}{4} (\bibinfo{year}{2017}),
  \bibinfo{pages}{98}.
\newblock


\bibitem[\protect\citeauthoryear{Chen, Liao, Yuan, Yu, and Hua}{Chen
  et~al\mbox{.}}{2017}]%
        {chen2017coherent}
\bibfield{author}{\bibinfo{person}{Dongdong Chen}, \bibinfo{person}{Jing Liao},
  \bibinfo{person}{Lu Yuan}, \bibinfo{person}{Nenghai Yu}, {and}
  \bibinfo{person}{Gang Hua}.} \bibinfo{year}{2017}\natexlab{}.
\newblock \showarticletitle{Coherent online video style transfer}. In
  \bibinfo{booktitle}{\emph{Proceedings of the IEEE International Conference on
  Computer Vision}}. \bibinfo{pages}{1105--1114}.
\newblock


\bibitem[\protect\citeauthoryear{Chu and Thuerey}{Chu and Thuerey}{2017}]%
        {chu2017data}
\bibfield{author}{\bibinfo{person}{Mengyu Chu} {and} \bibinfo{person}{Nils
  Thuerey}.} \bibinfo{year}{2017}\natexlab{}.
\newblock \showarticletitle{Data-driven synthesis of smoke flows with CNN-based
  feature descriptors}.
\newblock \bibinfo{journal}{\emph{ACM Transactions on Graphics (TOG)}}
  \bibinfo{volume}{36}, \bibinfo{number}{4} (\bibinfo{year}{2017}),
  \bibinfo{pages}{69}.
\newblock


\bibitem[\protect\citeauthoryear{Chu, Xie, Leal-Taix{\'e}, and Thuerey}{Chu
  et~al\mbox{.}}{2018}]%
        {chu2018temporally}
\bibfield{author}{\bibinfo{person}{Mengyu Chu}, \bibinfo{person}{You Xie},
  \bibinfo{person}{Laura Leal-Taix{\'e}}, {and} \bibinfo{person}{Nils
  Thuerey}.} \bibinfo{year}{2018}\natexlab{}.
\newblock \showarticletitle{Temporally Coherent GANs for Video Super-Resolution
  (TecoGAN)}.
\newblock \bibinfo{journal}{\emph{arXiv preprint arXiv:1811.09393}}
  (\bibinfo{year}{2018}).
\newblock


\bibitem[\protect\citeauthoryear{Dong, Loy, He, and Tang}{Dong
  et~al\mbox{.}}{2014}]%
        {dong2014learning}
\bibfield{author}{\bibinfo{person}{Chao Dong}, \bibinfo{person}{Chen~Change
  Loy}, \bibinfo{person}{Kaiming He}, {and} \bibinfo{person}{Xiaoou Tang}.}
  \bibinfo{year}{2014}\natexlab{}.
\newblock \showarticletitle{Learning a deep convolutional network for image
  super-resolution}. In \bibinfo{booktitle}{\emph{European conference on
  computer vision}}. Springer, \bibinfo{pages}{184--199}.
\newblock


\bibitem[\protect\citeauthoryear{Farimani, Gomes, and Pande}{Farimani
  et~al\mbox{.}}{2017}]%
        {farimani2017deep}
\bibfield{author}{\bibinfo{person}{Amir~Barati Farimani},
  \bibinfo{person}{Joseph Gomes}, {and} \bibinfo{person}{Vijay~S Pande}.}
  \bibinfo{year}{2017}\natexlab{}.
\newblock \showarticletitle{Deep learning the physics of transport phenomena}.
\newblock \bibinfo{journal}{\emph{arXiv preprint arXiv:1709.02432}}
  (\bibinfo{year}{2017}).
\newblock


\bibitem[\protect\citeauthoryear{Goodfellow, Pouget-Abadie, Mirza, Xu,
  Warde-Farley, Ozair, Courville, and Bengio}{Goodfellow et~al\mbox{.}}{2014}]%
        {goodfellow2014generative}
\bibfield{author}{\bibinfo{person}{Ian Goodfellow}, \bibinfo{person}{Jean
  Pouget-Abadie}, \bibinfo{person}{Mehdi Mirza}, \bibinfo{person}{Bing Xu},
  \bibinfo{person}{David Warde-Farley}, \bibinfo{person}{Sherjil Ozair},
  \bibinfo{person}{Aaron Courville}, {and} \bibinfo{person}{Yoshua Bengio}.}
  \bibinfo{year}{2014}\natexlab{}.
\newblock \showarticletitle{Generative adversarial nets}. In
  \bibinfo{booktitle}{\emph{Advances in neural information processing
  systems}}. \bibinfo{pages}{2672--2680}.
\newblock


\bibitem[\protect\citeauthoryear{He, Zhang, Ren, and Sun}{He
  et~al\mbox{.}}{2016}]%
        {he2016identity}
\bibfield{author}{\bibinfo{person}{Kaiming He}, \bibinfo{person}{Xiangyu
  Zhang}, \bibinfo{person}{Shaoqing Ren}, {and} \bibinfo{person}{Jian Sun}.}
  \bibinfo{year}{2016}\natexlab{}.
\newblock \showarticletitle{Identity mappings in deep residual networks}. In
  \bibinfo{booktitle}{\emph{European conference on computer vision}}. Springer,
  \bibinfo{pages}{630--645}.
\newblock


\bibitem[\protect\citeauthoryear{Jeong, Solenthaler, Pollefeys, Gross,
  et~al\mbox{.}}{Jeong et~al\mbox{.}}{2015}]%
        {jeong2015data}
\bibfield{author}{\bibinfo{person}{SoHyeon Jeong}, \bibinfo{person}{Barbara
  Solenthaler}, \bibinfo{person}{Marc Pollefeys}, \bibinfo{person}{Markus
  Gross}, {et~al\mbox{.}}} \bibinfo{year}{2015}\natexlab{}.
\newblock \showarticletitle{Data-driven fluid simulations using regression
  forests}.
\newblock \bibinfo{journal}{\emph{ACM Transactions on Graphics (TOG)}}
  \bibinfo{volume}{34}, \bibinfo{number}{6} (\bibinfo{year}{2015}),
  \bibinfo{pages}{199}.
\newblock


\bibitem[\protect\citeauthoryear{Johnson, Alahi, and Fei-Fei}{Johnson
  et~al\mbox{.}}{2016}]%
        {johnson2016perceptual}
\bibfield{author}{\bibinfo{person}{Justin Johnson}, \bibinfo{person}{Alexandre
  Alahi}, {and} \bibinfo{person}{Li Fei-Fei}.} \bibinfo{year}{2016}\natexlab{}.
\newblock \showarticletitle{Perceptual losses for real-time style transfer and
  super-resolution}. In \bibinfo{booktitle}{\emph{European conference on
  computer vision}}. Springer, \bibinfo{pages}{694--711}.
\newblock


\bibitem[\protect\citeauthoryear{Kallweit, M{\"u}ller, McWilliams, Gross, and
  Nov{\'a}k}{Kallweit et~al\mbox{.}}{2017}]%
        {kallweit2017deep}
\bibfield{author}{\bibinfo{person}{Simon Kallweit}, \bibinfo{person}{Thomas
  M{\"u}ller}, \bibinfo{person}{Brian McWilliams}, \bibinfo{person}{Markus
  Gross}, {and} \bibinfo{person}{Jan Nov{\'a}k}.}
  \bibinfo{year}{2017}\natexlab{}.
\newblock \showarticletitle{Deep scattering: Rendering atmospheric clouds with
  radiance-predicting neural networks}.
\newblock \bibinfo{journal}{\emph{ACM Transactions on Graphics (TOG)}}
  \bibinfo{volume}{36}, \bibinfo{number}{6} (\bibinfo{year}{2017}),
  \bibinfo{pages}{231}.
\newblock


\bibitem[\protect\citeauthoryear{Karras, Aila, Laine, and Lehtinen}{Karras
  et~al\mbox{.}}{2017}]%
        {karras2017progressive}
\bibfield{author}{\bibinfo{person}{Tero Karras}, \bibinfo{person}{Timo Aila},
  \bibinfo{person}{Samuli Laine}, {and} \bibinfo{person}{Jaakko Lehtinen}.}
  \bibinfo{year}{2017}\natexlab{}.
\newblock \showarticletitle{Progressive growing of gans for improved quality,
  stability, and variation}.
\newblock \bibinfo{journal}{\emph{arXiv preprint arXiv:1710.10196}}
  (\bibinfo{year}{2017}).
\newblock


\bibitem[\protect\citeauthoryear{Kennedy, Israel, Frenkel, Bar-Shalom, and
  Azhari}{Kennedy et~al\mbox{.}}{2006}]%
        {kennedy2006super}
\bibfield{author}{\bibinfo{person}{John~A Kennedy}, \bibinfo{person}{Ora
  Israel}, \bibinfo{person}{Alex Frenkel}, \bibinfo{person}{Rachel Bar-Shalom},
  {and} \bibinfo{person}{Haim Azhari}.} \bibinfo{year}{2006}\natexlab{}.
\newblock \showarticletitle{Super-resolution in PET imaging}.
\newblock \bibinfo{journal}{\emph{IEEE transactions on medical imaging}}
  \bibinfo{volume}{25}, \bibinfo{number}{2} (\bibinfo{year}{2006}),
  \bibinfo{pages}{137--147}.
\newblock


\bibitem[\protect\citeauthoryear{Kim, Azevedo, Thuerey, Kim, Gross, and
  Solenthaler}{Kim et~al\mbox{.}}{2018}]%
        {kim2018deep}
\bibfield{author}{\bibinfo{person}{Byungsoo Kim}, \bibinfo{person}{Vinicius~C
  Azevedo}, \bibinfo{person}{Nils Thuerey}, \bibinfo{person}{Theodore Kim},
  \bibinfo{person}{Markus Gross}, {and} \bibinfo{person}{Barbara Solenthaler}.}
  \bibinfo{year}{2018}\natexlab{}.
\newblock \showarticletitle{Deep Fluids: A Generative Network for Parameterized
  Fluid Simulations}.
\newblock \bibinfo{journal}{\emph{arXiv preprint arXiv:1806.02071}}
  (\bibinfo{year}{2018}).
\newblock


\bibitem[\protect\citeauthoryear{Kim, Liu, Llamas, and Rossignac}{Kim
  et~al\mbox{.}}{2005}]%
        {kim2005flowfixer}
\bibfield{author}{\bibinfo{person}{ByungMoon Kim}, \bibinfo{person}{Yingjie
  Liu}, \bibinfo{person}{Ignacio Llamas}, {and} \bibinfo{person}{Jaroslaw~R
  Rossignac}.} \bibinfo{year}{2005}\natexlab{}.
\newblock \bibinfo{booktitle}{\emph{Flowfixer: Using bfecc for fluid
  simulation}}.
\newblock \bibinfo{type}{{T}echnical {R}eport}. \bibinfo{institution}{Georgia
  Institute of Technology}.
\newblock


\bibitem[\protect\citeauthoryear{Kim, Kwon~Lee, and Mu~Lee}{Kim
  et~al\mbox{.}}{2016}]%
        {kim2016accurate}
\bibfield{author}{\bibinfo{person}{Jiwon Kim}, \bibinfo{person}{Jung Kwon~Lee},
  {and} \bibinfo{person}{Kyoung Mu~Lee}.} \bibinfo{year}{2016}\natexlab{}.
\newblock \showarticletitle{Accurate image super-resolution using very deep
  convolutional networks}. In \bibinfo{booktitle}{\emph{Proceedings of the IEEE
  conference on computer vision and pattern recognition}}.
  \bibinfo{pages}{1646--1654}.
\newblock


\bibitem[\protect\citeauthoryear{Kim, Th{\"u}rey, James, and Gross}{Kim
  et~al\mbox{.}}{2008}]%
        {kim2008wavelet}
\bibfield{author}{\bibinfo{person}{Theodore Kim}, \bibinfo{person}{Nils
  Th{\"u}rey}, \bibinfo{person}{Doug James}, {and} \bibinfo{person}{Markus
  Gross}.} \bibinfo{year}{2008}\natexlab{}.
\newblock \showarticletitle{Wavelet turbulence for fluid simulation}. In
  \bibinfo{booktitle}{\emph{ACM Transactions on Graphics (TOG)}},
  Vol.~\bibinfo{volume}{27}. ACM, \bibinfo{pages}{50}.
\newblock


\bibitem[\protect\citeauthoryear{Kingma and Ba}{Kingma and Ba}{2014}]%
        {Adam}
\bibfield{author}{\bibinfo{person}{Diederik~P. Kingma} {and}
  \bibinfo{person}{Jimmy Ba}.} \bibinfo{year}{2014}\natexlab{}.
\newblock \showarticletitle{Adam: {A} Method for Stochastic Optimization}.
\newblock \bibinfo{journal}{\emph{CoRR}}  \bibinfo{volume}{abs/1412.6980}
  (\bibinfo{year}{2014}).
\newblock
\showeprint[arxiv]{1412.6980}
\urldef\tempurl%
\url{http://arxiv.org/abs/1412.6980}
\showURL{%
\tempurl}


\bibitem[\protect\citeauthoryear{Lai, Huang, Ahuja, and Yang}{Lai
  et~al\mbox{.}}{2017}]%
        {lai2017deep}
\bibfield{author}{\bibinfo{person}{Wei-Sheng Lai}, \bibinfo{person}{Jia-Bin
  Huang}, \bibinfo{person}{Narendra Ahuja}, {and} \bibinfo{person}{Ming-Hsuan
  Yang}.} \bibinfo{year}{2017}\natexlab{}.
\newblock \showarticletitle{Deep laplacian pyramid networks for fast and
  accurate superresolution}. In \bibinfo{booktitle}{\emph{IEEE Conference on
  Computer Vision and Pattern Recognition}}, Vol.~\bibinfo{volume}{2}.
  \bibinfo{pages}{5}.
\newblock


\bibitem[\protect\citeauthoryear{Ledig, Theis, Husz{\'a}r, Caballero,
  Cunningham, Acosta, Aitken, Tejani, Totz, Wang, et~al\mbox{.}}{Ledig
  et~al\mbox{.}}{2017}]%
        {ledig2017photo}
\bibfield{author}{\bibinfo{person}{Christian Ledig}, \bibinfo{person}{Lucas
  Theis}, \bibinfo{person}{Ferenc Husz{\'a}r}, \bibinfo{person}{Jose
  Caballero}, \bibinfo{person}{Andrew Cunningham}, \bibinfo{person}{Alejandro
  Acosta}, \bibinfo{person}{Andrew Aitken}, \bibinfo{person}{Alykhan Tejani},
  \bibinfo{person}{Johannes Totz}, \bibinfo{person}{Zehan Wang},
  {et~al\mbox{.}}} \bibinfo{year}{2017}\natexlab{}.
\newblock \showarticletitle{Photo-realistic single image super-resolution using
  a generative adversarial network}.
\newblock \bibinfo{journal}{\emph{arXiv preprint}} (\bibinfo{year}{2017}).
\newblock


\bibitem[\protect\citeauthoryear{Lim, Son, Kim, Nah, and Lee}{Lim
  et~al\mbox{.}}{2017}]%
        {lim2017enhanced}
\bibfield{author}{\bibinfo{person}{Bee Lim}, \bibinfo{person}{Sanghyun Son},
  \bibinfo{person}{Heewon Kim}, \bibinfo{person}{Seungjun Nah}, {and}
  \bibinfo{person}{Kyoung~Mu Lee}.} \bibinfo{year}{2017}\natexlab{}.
\newblock \showarticletitle{Enhanced deep residual networks for single image
  super-resolution}. In \bibinfo{booktitle}{\emph{CVPR}},
  Vol.~\bibinfo{volume}{1}. \bibinfo{pages}{3}.
\newblock


\bibitem[\protect\citeauthoryear{Long, Lu, Ma, and Dong}{Long
  et~al\mbox{.}}{2017}]%
        {long2017pde}
\bibfield{author}{\bibinfo{person}{Zichao Long}, \bibinfo{person}{Yiping Lu},
  \bibinfo{person}{Xianzhong Ma}, {and} \bibinfo{person}{Bin Dong}.}
  \bibinfo{year}{2017}\natexlab{}.
\newblock \showarticletitle{Pde-net: Learning pdes from data}.
\newblock \bibinfo{journal}{\emph{arXiv preprint arXiv:1710.09668}}
  (\bibinfo{year}{2017}).
\newblock


\bibitem[\protect\citeauthoryear{Mao, Li, Xie, Lau, and Wang}{Mao
  et~al\mbox{.}}{2016}]%
        {LSGAN}
\bibfield{author}{\bibinfo{person}{Xudong Mao}, \bibinfo{person}{Qing Li},
  \bibinfo{person}{Haoran Xie}, \bibinfo{person}{Raymond Y.~K. Lau}, {and}
  \bibinfo{person}{Zhen Wang}.} \bibinfo{year}{2016}\natexlab{}.
\newblock \showarticletitle{Multi-class Generative Adversarial Networks with
  the {L2} Loss Function}.
\newblock \bibinfo{journal}{\emph{CoRR}}  \bibinfo{volume}{abs/1611.04076}
  (\bibinfo{year}{2016}).
\newblock
\showeprint[arxiv]{1611.04076}
\urldef\tempurl%
\url{http://arxiv.org/abs/1611.04076}
\showURL{%
\tempurl}


\bibitem[\protect\citeauthoryear{Mirza and Osindero}{Mirza and
  Osindero}{2014}]%
        {mirza2014conditional}
\bibfield{author}{\bibinfo{person}{Mehdi Mirza} {and} \bibinfo{person}{Simon
  Osindero}.} \bibinfo{year}{2014}\natexlab{}.
\newblock \showarticletitle{Conditional generative adversarial nets}.
\newblock \bibinfo{journal}{\emph{arXiv preprint arXiv:1411.1784}}
  (\bibinfo{year}{2014}).
\newblock


\bibitem[\protect\citeauthoryear{Narain, Sewall, Carlson, and Lin}{Narain
  et~al\mbox{.}}{2008}]%
        {narain2008fast}
\bibfield{author}{\bibinfo{person}{Rahul Narain}, \bibinfo{person}{Jason
  Sewall}, \bibinfo{person}{Mark Carlson}, {and} \bibinfo{person}{Ming~C Lin}.}
  \bibinfo{year}{2008}\natexlab{}.
\newblock \showarticletitle{Fast animation of turbulence using energy transport
  and procedural synthesis}. In \bibinfo{booktitle}{\emph{ACM Transactions on
  Graphics (TOG)}}, Vol.~\bibinfo{volume}{27}. ACM, \bibinfo{pages}{166}.
\newblock


\bibitem[\protect\citeauthoryear{Peng, Berseth, Yin, and Van De~Panne}{Peng
  et~al\mbox{.}}{2017}]%
        {peng2017deeploco}
\bibfield{author}{\bibinfo{person}{Xue~Bin Peng}, \bibinfo{person}{Glen
  Berseth}, \bibinfo{person}{KangKang Yin}, {and} \bibinfo{person}{Michiel Van
  De~Panne}.} \bibinfo{year}{2017}\natexlab{}.
\newblock \showarticletitle{Deeploco: Dynamic locomotion skills using
  hierarchical deep reinforcement learning}.
\newblock \bibinfo{journal}{\emph{ACM Transactions on Graphics (TOG)}}
  \bibinfo{volume}{36}, \bibinfo{number}{4} (\bibinfo{year}{2017}),
  \bibinfo{pages}{41}.
\newblock


\bibitem[\protect\citeauthoryear{Prantl, Bonev, and Thuerey}{Prantl
  et~al\mbox{.}}{2017}]%
        {prantl2017pre}
\bibfield{author}{\bibinfo{person}{Lukas Prantl}, \bibinfo{person}{Boris
  Bonev}, {and} \bibinfo{person}{Nils Thuerey}.}
  \bibinfo{year}{2017}\natexlab{}.
\newblock \showarticletitle{Pre-computed liquid spaces with generative neural
  networks and optical flow}.
\newblock \bibinfo{journal}{\emph{arXiv preprint arXiv:1704.07854}}
  (\bibinfo{year}{2017}).
\newblock


\bibitem[\protect\citeauthoryear{Radford, Metz, and Chintala}{Radford
  et~al\mbox{.}}{2016}]%
        {RadfordMC15}
\bibfield{author}{\bibinfo{person}{Alec Radford}, \bibinfo{person}{Luke Metz},
  {and} \bibinfo{person}{Soumith Chintala}.} \bibinfo{year}{2016}\natexlab{}.
\newblock \showarticletitle{Unsupervised Representation Learning with Deep
  Convolutional Generative Adversarial Networks}.
\newblock \bibinfo{journal}{\emph{Proc. ICLR}} (\bibinfo{year}{2016}).
\newblock


\bibitem[\protect\citeauthoryear{Ruder, Dosovitskiy, and Brox}{Ruder
  et~al\mbox{.}}{2016}]%
        {ruder2016artistic}
\bibfield{author}{\bibinfo{person}{Manuel Ruder}, \bibinfo{person}{Alexey
  Dosovitskiy}, {and} \bibinfo{person}{Thomas Brox}.}
  \bibinfo{year}{2016}\natexlab{}.
\newblock \showarticletitle{Artistic style transfer for videos}. In
  \bibinfo{booktitle}{\emph{German Conference on Pattern Recognition}}.
  Springer, \bibinfo{pages}{26--36}.
\newblock


\bibitem[\protect\citeauthoryear{Saito, Matsumoto, and Saito}{Saito
  et~al\mbox{.}}{2017}]%
        {saito2017temporal}
\bibfield{author}{\bibinfo{person}{Masaki Saito}, \bibinfo{person}{Eiichi
  Matsumoto}, {and} \bibinfo{person}{Shunta Saito}.}
  \bibinfo{year}{2017}\natexlab{}.
\newblock \showarticletitle{Temporal generative adversarial nets with singular
  value clipping}. In \bibinfo{booktitle}{\emph{Proceedings of the IEEE
  International Conference on Computer Vision}}. \bibinfo{pages}{2830--2839}.
\newblock


\bibitem[\protect\citeauthoryear{Sajjadi, Sch{\"o}lkopf, and Hirsch}{Sajjadi
  et~al\mbox{.}}{2017}]%
        {sajjadi2017enhancenet}
\bibfield{author}{\bibinfo{person}{Mehdi~SM Sajjadi}, \bibinfo{person}{Bernhard
  Sch{\"o}lkopf}, {and} \bibinfo{person}{Michael Hirsch}.}
  \bibinfo{year}{2017}\natexlab{}.
\newblock \showarticletitle{Enhancenet: Single image super-resolution through
  automated texture synthesis}. In \bibinfo{booktitle}{\emph{Computer Vision
  (ICCV), 2017 IEEE International Conference on}}. IEEE,
  \bibinfo{pages}{4501--4510}.
\newblock


\bibitem[\protect\citeauthoryear{Selle, Fedkiw, Kim, Liu, and Rossignac}{Selle
  et~al\mbox{.}}{2008}]%
        {selle2008unconditionally}
\bibfield{author}{\bibinfo{person}{Andrew Selle}, \bibinfo{person}{Ronald
  Fedkiw}, \bibinfo{person}{Byungmoon Kim}, \bibinfo{person}{Yingjie Liu},
  {and} \bibinfo{person}{Jarek Rossignac}.} \bibinfo{year}{2008}\natexlab{}.
\newblock \showarticletitle{An unconditionally stable MacCormack method}.
\newblock \bibinfo{journal}{\emph{Journal of Scientific Computing}}
  \bibinfo{volume}{35}, \bibinfo{number}{2-3} (\bibinfo{year}{2008}),
  \bibinfo{pages}{350--371}.
\newblock


\bibitem[\protect\citeauthoryear{Stam}{Stam}{1999}]%
        {stam1999stable}
\bibfield{author}{\bibinfo{person}{Jos Stam}.} \bibinfo{year}{1999}\natexlab{}.
\newblock \showarticletitle{Stable Fluids.}. In
  \bibinfo{booktitle}{\emph{Siggraph}}, Vol.~\bibinfo{volume}{99}.
  \bibinfo{pages}{121--128}.
\newblock


\bibitem[\protect\citeauthoryear{Tai, Yang, and Liu}{Tai et~al\mbox{.}}{2017}]%
        {tai2017image}
\bibfield{author}{\bibinfo{person}{Ying Tai}, \bibinfo{person}{Jian Yang},
  {and} \bibinfo{person}{Xiaoming Liu}.} \bibinfo{year}{2017}\natexlab{}.
\newblock \showarticletitle{Image super-resolution via deep recursive residual
  network}. In \bibinfo{booktitle}{\emph{Proceedings of the IEEE Conference on
  Computer Vision and Pattern Recognition}}, Vol.~\bibinfo{volume}{1}.
  \bibinfo{pages}{5}.
\newblock


\bibitem[\protect\citeauthoryear{Teng, Levin, and Kim}{Teng
  et~al\mbox{.}}{2016}]%
        {teng2016eulerian}
\bibfield{author}{\bibinfo{person}{Yun Teng}, \bibinfo{person}{David~IW Levin},
  {and} \bibinfo{person}{Theodore Kim}.} \bibinfo{year}{2016}\natexlab{}.
\newblock \showarticletitle{Eulerian solid-fluid coupling}.
\newblock \bibinfo{journal}{\emph{ACM Transactions on Graphics (TOG)}}
  \bibinfo{volume}{35}, \bibinfo{number}{6} (\bibinfo{year}{2016}),
  \bibinfo{pages}{200}.
\newblock


\bibitem[\protect\citeauthoryear{Thuerey and Pfaff}{Thuerey and Pfaff}{2018}]%
        {mantaflow}
\bibfield{author}{\bibinfo{person}{Nils Thuerey} {and} \bibinfo{person}{Tobias
  Pfaff}.} \bibinfo{year}{2018}\natexlab{}.
\newblock \bibinfo{title}{{MantaFlow}}.
\newblock
\newblock
\newblock
\shownote{\emph{http://mantaflow.com}.}


\bibitem[\protect\citeauthoryear{Timofte, De~Smet, and Van~Gool}{Timofte
  et~al\mbox{.}}{2014}]%
        {timofte2014a+}
\bibfield{author}{\bibinfo{person}{Radu Timofte}, \bibinfo{person}{Vincent
  De~Smet}, {and} \bibinfo{person}{Luc Van~Gool}.}
  \bibinfo{year}{2014}\natexlab{}.
\newblock \showarticletitle{A+: Adjusted anchored neighborhood regression for
  fast super-resolution}. In \bibinfo{booktitle}{\emph{Asian conference on
  computer vision}}. Springer, \bibinfo{pages}{111--126}.
\newblock


\bibitem[\protect\citeauthoryear{Tompson, Schlachter, Sprechmann, and
  Perlin}{Tompson et~al\mbox{.}}{2017}]%
        {tompson2017accelerating}
\bibfield{author}{\bibinfo{person}{Jonathan Tompson},
  \bibinfo{person}{Kristofer Schlachter}, \bibinfo{person}{Pablo Sprechmann},
  {and} \bibinfo{person}{Ken Perlin}.} \bibinfo{year}{2017}\natexlab{}.
\newblock \showarticletitle{Accelerating eulerian fluid simulation with
  convolutional networks}. In \bibinfo{booktitle}{\emph{Proceedings of the 34th
  International Conference on Machine Learning-Volume 70}}. JMLR. org,
  \bibinfo{pages}{3424--3433}.
\newblock


\bibitem[\protect\citeauthoryear{Tong, Li, Liu, and Gao}{Tong
  et~al\mbox{.}}{2017}]%
        {tong2017image}
\bibfield{author}{\bibinfo{person}{Tong Tong}, \bibinfo{person}{Gen Li},
  \bibinfo{person}{Xiejie Liu}, {and} \bibinfo{person}{Qinquan Gao}.}
  \bibinfo{year}{2017}\natexlab{}.
\newblock \showarticletitle{Image super-resolution using dense skip
  connections}. In \bibinfo{booktitle}{\emph{Computer Vision (ICCV), 2017 IEEE
  International Conference on}}. IEEE, \bibinfo{pages}{4809--4817}.
\newblock


\bibitem[\protect\citeauthoryear{Um, Hu, and Thuerey}{Um et~al\mbox{.}}{2018}]%
        {um2018liquid}
\bibfield{author}{\bibinfo{person}{Kiwon Um}, \bibinfo{person}{Xiangyu Hu},
  {and} \bibinfo{person}{Nils Thuerey}.} \bibinfo{year}{2018}\natexlab{}.
\newblock \showarticletitle{Liquid splash modeling with neural networks}. In
  \bibinfo{booktitle}{\emph{Computer Graphics Forum}},
  Vol.~\bibinfo{volume}{37}. Wiley Online Library, \bibinfo{pages}{171--182}.
\newblock


\bibitem[\protect\citeauthoryear{Wang, Perazzi, McWilliams, Sorkine{-}Hornung,
  Sorkine{-}Hornung, and Schroers}{Wang et~al\mbox{.}}{2018}]%
        {wang2018progressive}
\bibfield{author}{\bibinfo{person}{Yifan Wang}, \bibinfo{person}{Federico
  Perazzi}, \bibinfo{person}{Brian McWilliams}, \bibinfo{person}{Alexander
  Sorkine{-}Hornung}, \bibinfo{person}{Olga Sorkine{-}Hornung}, {and}
  \bibinfo{person}{Christopher Schroers}.} \bibinfo{year}{2018}\natexlab{}.
\newblock \showarticletitle{A Fully Progressive Approach to Single-Image
  Super-Resolution}.
\newblock \bibinfo{journal}{\emph{CoRR}}  \bibinfo{volume}{abs/1804.02900}
  (\bibinfo{year}{2018}).
\newblock
\showeprint[arxiv]{1804.02900}
\urldef\tempurl%
\url{http://arxiv.org/abs/1804.02900}
\showURL{%
\tempurl}


\bibitem[\protect\citeauthoryear{Xie, Franz, Chu, and Thuerey}{Xie
  et~al\mbox{.}}{2018}]%
        {xie2018tempogan}
\bibfield{author}{\bibinfo{person}{You Xie}, \bibinfo{person}{Erik Franz},
  \bibinfo{person}{Mengyu Chu}, {and} \bibinfo{person}{Nils Thuerey}.}
  \bibinfo{year}{2018}\natexlab{}.
\newblock \showarticletitle{tempoGAN: A Temporally Coherent, Volumetric GAN for
  Super-resolution Fluid Flow}.
\newblock \bibinfo{journal}{\emph{arXiv preprint arXiv:1801.09710}}
  (\bibinfo{year}{2018}).
\newblock


\bibitem[\protect\citeauthoryear{Yu, Zhang, Wang, and Yu}{Yu
  et~al\mbox{.}}{2017}]%
        {yu2017seqgan}
\bibfield{author}{\bibinfo{person}{Lantao Yu}, \bibinfo{person}{Weinan Zhang},
  \bibinfo{person}{Jun Wang}, {and} \bibinfo{person}{Yong Yu}.}
  \bibinfo{year}{2017}\natexlab{}.
\newblock \showarticletitle{Seqgan: Sequence generative adversarial nets with
  policy gradient}. In \bibinfo{booktitle}{\emph{Thirty-First AAAI Conference
  on Artificial Intelligence}}.
\newblock


\bibitem[\protect\citeauthoryear{Zhang, Zhang, Shen, and Li}{Zhang
  et~al\mbox{.}}{2010}]%
        {zhang2010super}
\bibfield{author}{\bibinfo{person}{Liangpei Zhang}, \bibinfo{person}{Hongyan
  Zhang}, \bibinfo{person}{Huanfeng Shen}, {and} \bibinfo{person}{Pingxiang
  Li}.} \bibinfo{year}{2010}\natexlab{}.
\newblock \showarticletitle{A super-resolution reconstruction algorithm for
  surveillance images}.
\newblock \bibinfo{journal}{\emph{Signal Processing}} \bibinfo{volume}{90},
  \bibinfo{number}{3} (\bibinfo{year}{2010}), \bibinfo{pages}{848--859}.
\newblock


\end{thebibliography}

\end{document}